\documentclass[paper]{JHEP3}

\usepackage{epsfig,multicol}

\def\a{\alpha}
\def\b{\beta}
\def\c{\gamma}
\def\d{\delta}
\def\e{\epsilon}

\def\l{\lambda}
\def\m{\mu}
\def\n{\nu}

\def\r{\rho}
\def\s{\sigma}

\def\w{\omega}

\def\C{\Gamma}
\def\D{\Delta}
\def\L{\Lambda}

\def\W{\Omega}

\def\ul{\underline}
\def\rta{\rightarrow}
\def\orta{\overrightarrow}

\def\AA{{\cal A}}
\def\BB{{\cal B}}
\def\LL{{\cal L}}
\def\FF{{\cal F}}
\def\GG{{\cal G}}
\def\XX{{\cal X}}

\def\RR{{\cal R}}
\def\MM{{\cal M}}
\def\tr{{\rm tr}}

\def\Dslash{\,{\raise.15ex\hbox{/}\mkern-12mu D}}

\def\Rea{{\rm Re~}}
\def\Ima{{\rm Im~}}

\title{Superluminality and UV Completion
\thanks{This research is supported in part by 
PPARC grant PP/D507407/1.  }}

\author{G.M. Shore\\
        Department of Physics\\
        University of Wales, Swansea\\
        Swansea SA2 8PP, U.K.\\
        E-mail: \email{g.m.shore@swansea.ac.uk}}

\abstract{The idea that the existence of a consistent UV completion 
satisfying the fundamental axioms of local quantum field theory or string
theory may impose positivity constraints on the couplings of the leading
irrelevant operators in a low-energy effective field theory is critically
discussed. Violation of these constraints implies superluminal propagation,
in the sense that the low-frequency limit of the phase velocity  
$v_{\rm ph}(0)$ exceeds $c$. It is explained why causality is related not to
$v_{\rm ph}(0)$ but to the high-frequency limit $v_{\rm ph}(\infty)$ and
how these are related by the Kramers-Kronig dispersion relation, depending
on the sign of the imaginary part of the refractive index $\Ima n(\w)$
which is normally assumed positive. Superluminal propagation and its relation 
to UV completion is investigated in detail in three theories: QED in a 
background electromagnetic field, where the full dispersion relation for
$n(\w)$ is evaluated numerically for the first time and the role of the 
null energy condition $T_{\m\n}k^\m k^\n \ge 0$ is highlighted;
QED in a background gravitational field, where examples of superluminal
low-frequency phase velocities arise in violation of the positivity
constraints; and light propagation in coupled laser-atom $\L$-systems
exhibiting Raman gain lines with $\Ima n(\w) < 0$. The possibility that 
a negative $\Ima n(\w)$ must occur in quantum field theories involving 
gravity to avoid causality violation, and the implications for the 
relation of IR effective field theories to their UV completion, are 
carefully analysed.
}

\begin{document}

\section{Introduction}

Effective field theories have always played an important role in particle 
physics as low-energy phenomenological models describing the IR dynamics
of renormalisable quantum field theories \cite{Weinberg:1978kz}. Key examples
include the chiral Lagrangians describing the light pseudoscalar mesons
arising from chiral symmetry breaking in QCD or those related to
electroweak symmetry breaking in the absence of a light Higgs boson.
More recently, the necessity of renormalisability as a criterion in
constructing unified field theories has been reconsidered and it has become
familiar to think of the standard model and its supersymmetric 
generalisations as IR effective theories of some more fundamental QFT or 
string theory at the Planck scale. 

The question then arises whether {\it all} effective field theories 
admit a consistent UV completion, i.e.~whether they can arise as the 
IR limit of a well-defined renormalisable QFT or string theory. If not,
we may be able to use the criterion of the existence of a UV completion 
as a constraint on the parameters of the IR effective field theory, 
perhaps with interesting phenomenological implications.

At first sight, the answer certainly appears to be no. It is well-known 
that the parameters of chiral Lagrangians derived from
known renormalisable theories such as QCD satisfy constraints arising
from analyticity and unitarity. However, in general the situation may not 
be so straightforward, particularly when we consider theories involving
gravity. The purpose of this paper is to expose some of the subtleties
that arise in establishing the relation between constraints on IR
effective field theories and their UV completion, especially those
that arise from considerations of causality and the absence of
superluminal propagation.

The idea that the existence of a well-defined UV completion may place
important constraints on the couplings of the leading irrelevant operators
in low-energy effective actions has been revisited recently in  
ref.\cite{Adams:2006sv}.
This paper highlights a number of examples where these couplings must 
satisfy certain positivity constraints to ensure that the effective theory
does not exhibit unphysical behaviour such as superluminal propagation
of massless particles or violations of analyticity or unitarity bounds
on scattering amplitudes. This has been developed in 
ref.\cite{Distler:2006if}, where positivity constraints are derived on
the couplings of the operators controlling WW scattering in
an effective theory of electroweak symmetry breaking. Turning the argument
around, these papers claim that an observed violation of these positivity
constraints would signify a breakdown of some of the fundamental axioms of
QFT and string theory, such as Lorentz invariance, unitarity,
analyticity or causality.

In this paper, we examine these arguments critically and show that
the assumed relation between the IR effective field theory and its
UV completion may be significantly more subtle. We focus on the issue of
superluminal propagation in the effective theory. This makes contact with
our extensive body of work (see e.g.~\cite{Drummond:1979pp,Shore:1995fz,
Shore:2002gn,Shore:2002gw,Shore:2003jx,Shore:2003zc,Shore:2004sh})
on superluminality, causality and dispersion in QED in curved spacetime,
a subject we have given the name of `quantum gravitational optics'
\cite{Shore:2003zc}. As will become clear, our discussion is equally
applicable to the issue of analyticity bounds, since each ultimately 
rests on the behaviour of the imaginary (absorptive) part of the
refractive index or forward scattering amplitude.
 
We begin by discussing the general form of the effective action for 
low-energy electrodynamics and derive positivity constraints on the
couplings of the leading irrelevant operators based on the requirement
that light propagation is not superluminal. However, the `speed of light'
obtained from the IR effective theory is simply the zero-frequency 
limit of the phase velocity, $v_{\rm ph}(0)$. In section 3, we present 
a careful analysis of dispersion and light propagation and show that
in fact the `speed of light' relevant for causality is $v_{\rm ph}(\infty)$,
i.e.~the {\it high-frequency} limit of the phase velocity. Determining this
requires a knowledge of the UV completion of the quantum field theory.

It therefore appears that causality and superluminal propagation place 
no restriction on the low-energy theory. However, on the basis of the
usual axioms of quantum field theory (local Lorentz invariance,
analyticity, microcausality) we can prove the Kramers-Kronig dispersion
relation for the refractive index:
\begin{equation}
n(\infty) ~=~ n(0) ~-~ {2\over\pi} \int_0^\infty {d\w\over\w}~ \Ima n(\w) 
\label{eq:aa}
\end{equation}
It is usually assumed that $\Ima n(\w) > 0$, typical of an absorptive medium.
(Similar relations hold for scattering amplitudes. Here, the imaginary
part of the forward scattering amplitude, $\Ima \MM(s,0)$, is related
via the optical theorem to the total cross section, ensuring positivity.)
If this is true, then eq.(\ref{eq:aa}) implies that 
$v_{\rm ph}(\infty) > v_{\rm ph}(0)$. It follows that if $v_{\rm ph}(0)$
is found to be superluminal in the low-energy theory then $v_{\rm ph}(\infty)$
is necessarily also superluminal, in contradiction with causality. 
Violation of the positivity constraints on the couplings in the IR effective
action would indeed then mean that the theory had no consistent UV completion.
The central point of this paper, however, is to question the assumption
that $\Ima n(\w)$ is necessarily positive, especially in theories 
involving gravity.

First, we consider an example that is under complete control and where
the conventional expectations are realised, viz.~QED in a background
electromagnetic field. The background field modifies the propagation of 
light due to vacuum polarisation and induces irrelevant operators in the
low-energy effective action. Recent technical developments mean that we 
now have an expression for the vacuum polarisation valid for all momenta
for general constant background fields \cite{Schubert:2000yt}
and in section 4 we use this to derive the full frequency dependence of the
refractive index $n(\w)$. This is the first complete numerical evaluation
of the QED dispersion relation. This shows the expected features -- a single
absorption line with $\Ima n(\w) > 0$ and the corresponding phase
velocity satisfying $v_{\rm ph}(\infty) > v_{\rm ph}(0)$. The low-energy
couplings, which determine the Euler-Heisenberg effective action, are
consistent with the positivity constraints following from imposing
$v_{\rm ph}(0) < 1$.\footnote{Note that in general we set $c=1$ throughout
this paper.}

An important feature of our analysis is the demonstration of the precise role
of the {\it null energy condition} $T_{\m\n}k^\m k^\n \ge 0$, where $k^\m$
is the photon momentum. 
(See also, e.g., refs.~\cite{Shore:1995fz,Visser:1998ua,Dubovsky:2005xd}.)
The null projection of the energy-momentum tensor 
occurs universally in the dispersion relation for light propagation in
background fields -- the speed of light is determined by {\it both}
the couplings {\it and} the sign of $T_{\m\n}k^\m k^\n$. This is greatly
clarified by our use of the Newman-Penrose, or null-tetrad, formalism,
which is by far the clearest language in which to analyse the propagation
of light in general backgrounds.

We then consider the analogous case of QED in a background gravitational
field, i.e.~in curved spacetime. Here, current techniques only allow
a determination of the refractive index and phase velocity in the 
low-frequency region. Moreover, the null-energy projection 
$T_{\m\n}k^\m k^\n$ (the Newman-Penrose scalar $\Phi_{00}$) is not the
only quantity determining the speed of light. There is also a 
polarisation-dependent contribution given by the Weyl tensor projection
$C_{\m\l\n\r}k^\l k^\r a^\m a^\n$ (essentially the NP scalar $\Psi_0$), 
which gives rise to the phenomenon of gravitational birefringence. 
The remarkable feature of QED in curved spacetime is that, even for Weyl-flat
spacetimes, the positivity constraints on the low-energy couplings which
would prohibit superluminal propagation are violated \cite{Drummond:1979pp}. 
We can readily find examples exhibiting a superluminal low-frequency phase 
velocity, $v_{\rm ph}(0) > 1$. Furthermore, for Ricci-flat spacetimes 
it is possible to prove a simple sum rule \cite{Shore:1995fz} which shows 
that one polarisation is always superluminal, whatever the sign of the 
couplings. We illustrate these phenomena with examples of propagation in
FRW and black-hole spacetimes in section 5.

This raises the central issue of this paper. If the KK dispersion relation
holds in gravitational backgrounds and $\Ima n(\w) > 0$, then necessarily
$v_{\rm ph}(\infty) > v_{\rm ph}(0) > 1$ for certain spacetimes, in 
apparent violation of causality. (See ref.\cite{Shore:2003jx} and sections
3 and 5 for a more nuanced discussion).
The most natural escape seems to be that for theories involving gravity
we must have the possibility that $\Ima n(\w) < 0$. But if so, this 
invalidates the conclusion that IR effective field theories with couplings
violating the positivity constraints {\it necessarily} have no consistent
UV completion. Once we admit the possibility that a fundamental QFT 
satisfying the standard axioms (and surely QED in curved spacetime is such 
a theory) can have a negative absorptive part, $\Ima n(\w) < 0$ or
$\Ima \MM(s,0) < 0$, then we must accept that the relation between
the IR effective theory and its UV completion can be rather more subtle.

In order to understand better what a negative $\Ima n(\w)$ would mean 
physically, in section 6 we introduce a fascinating topic in atomic physics --
the interaction of lasers with so-called `$\L$-systems' exhibiting 
electromagnetically-induced transparency (EIT) and `slow light'. We discuss
a variant in which the $\L$-system is engineered to produce Raman {\it gain} 
lines \cite{Wang} rather than absorption lines in $\Ima n(\w)$ and derive in 
detail an explicit example where $v_{\rm ph}(\infty) < v_{\rm ph}(0)$. This 
occurs as a result of the intensity gain of a light pulse propagating through 
the coupled laser-atom medium. Our conjecture is that gravity may be able to 
act in the same way.

The issue of whether the existence of a consistent UV completion does
indeed impose positivity constraints on the couplings of an IR effective
field theory therefore remains open. In particular, theories involving gravity
present a serious challenge to the accepted wisdom on the relation of
the IR and UV behaviour of QFT. A summary of our final conclusions is
presented in section 7.

\vfill\eject

\section{Superluminality constraints in low-energy quantum electrodynamics}

The low-energy effective action for QED, valid for momenta below the scale
of the electron mass $m$, is given by the well-known Euler-Heisenberg
Lagrangian,
\begin{equation}
\LL ~~=~~ -{1\over4}F_{\m\n}F^{\m\n} ~+~ {\a^2\over m^4}\Bigl(
-{1\over36} (F_{\m\n}F^{\m\n})^2 + {7\over90} F_{\m\n}F_{\l\r}
F^{\m\l}F^{\n\r} \Bigr)
\label{eq:ba}
\end{equation}
This is more conveniently written in terms of the two Lorentz scalars
quadratic in the field strengths, $\FF = {1\over4}F_{\m\n}F^{\m\n}$
and $\GG = {1\over4}F_{\m\n}\tilde F^{\m\n}$, as follows:
\begin{equation}
\LL ~~=~~ - \FF ~+~ {2\over45}{\a^2\over m^4}\Bigl(c_1 {\FF}^2
+ c_2 {\GG}^2 \Bigr)
\label{eq:bb}
\end{equation}
where $c_1 = 4$ and $c_2 =7$.

The higher-order terms omitted in eqs.(\ref{eq:ba}) or (\ref{eq:bb}) are of 
two types: higher orders in derivatives, i.e.~terms of $O(D^2 F^4/m^6)$
which give contributions to the photon dispersion relation 
suppressed by $O(k^2/m^2)$\footnote{Here, we are using `$k^2$' to
denote two powers of momentum, not necessarily the contracted form
$k^\m k_\m$ which can of course be zero on-shell even for large values
of $|{\underline k}|$. Similarly for `$D^2$.}, and higher orders in the
field strengths $\FF, \GG$ which are associated with higher
powers of $\a$. 

In this section, we consider photon propagation in a background 
electromagnetic field using this effective Lagrangian and discuss the
constraints imposed on the coefficients $c_1, c_2$ of the leading (dim 8) 
irrelevant operators in the low-energy expansion (\ref{eq:bb}) by the
requirement that propagation is causal, i.e.~subluminal.

The simplest way to study the causal nature of photon propagation is to
use geometric optics (see, e.g.~refs.\cite{Shore:2003jx,Shore:2003zc}
for reviews of our formalism). 
Since we shall be considering QED in a background
gravitational field later in this paper, the following brief summary of
geometric optics is sufficiently general to include the case of propagation
in an arbitrary spacetime metric $g_{\m\n}$. The electromagnetic field in the 
(modified) Maxwell equations derived from eq.(\ref{eq:bb}) is written in
terms of a slowly-varying amplitude (and polarisation) and a rapidly-varying
phase as follows:
\begin{equation}
\bigl(A_\m + i \e B_\m + \ldots \bigr) e^{i\vartheta/\e}
\label{eq:bc}  
\end{equation}
where $\e$ is a parameter introduced to keep track of the relative orders
of terms as the Maxwell and Bianchi equations are solved order-by-order
in $\e$. The wave vector is identified as the gradient of the phase, 
$k_\m = \partial_\m\vartheta$. We also write $A_\m = A a_\m$, where $A$ 
represents the amplitude itself while $a_\m$ specifies the polarisation.

Solving the usual Maxwell equation $D_\m F^{\m\n} = 0$, we find at $O(1/\e)$,
\begin{equation}
k^2 = 0
\label{eq:bd}
\end{equation}
while at $O(1)$,
\begin{equation}
k^\m D_\m a^\n = 0
\label{eq:be}
\end{equation}
and
\begin{equation}
k^\m D_\m(\ln A) = -{1\over2}D_\m k^\m
\label{eq:bf}
\end{equation}
Eq.(\ref{eq:bd}) shows immediately that $k^\m$ is a null vector. From its
definition as a gradient, we also see
\begin{equation}
k^\m D_\m k^\n  = k^\m D^\n k_\m = {1\over2}D^\n k^2 = 0
\label{eq:bg}
\end{equation}
Light rays, or equivalently photon trajectories, are the integral curves of 
$k^\m$, i.e.~the curves $x^\m(s)$ where $dx^\m/ds = k^\m$. These curves 
therefore satisfy
\begin{equation}
0 ~=~ k^\m D_\m k^\n 
~=~ {d^2 x^\n \over ds^2} + \C^\n_{\m\l} {dx^\m \over ds}{dx^\l \over ds}
\label{eq:bh}
\end{equation}
This is the geodesic equation. We conclude that for the usual Maxwell theory
in general relativity, light follows null geodesics. 
Eqs.(\ref{eq:bg}),(\ref{eq:be}) show that both the wave vector and the 
polarisation are parallel transported along these null geodesic rays, 
while eq.(\ref{eq:bf}), whose r.h.s.~is just (minus) the optical scalar 
$\theta$, shows how the amplitude changes as the beam of rays focuses 
or diverges.

As a consequence of only keeping terms in the effective action 
with no explicit derivatives acting on the field strengths, the modified
light-cone condition derived from the low-energy effective action (\ref{eq:bb})
remains homogeneous and quadratic in $k^\m$. It can therefore be written 
in the form:
\begin{equation}
{\cal G}^{\m\n}k_\m k_\n = 0
\label{eq:bj}
\end{equation}
where ${\cal G}^{\m\n}$ is a function of the background
electromagnetic (or gravitational) field. In the gravitational case, 
${\cal G}^{\m\n}$ also involves the photon polarisation.

Now notice that in the discussion of the free Maxwell theory, we did not need 
to distinguish between the photon momentum  $p^\m$, i.e.~the tangent vector 
to the light rays, and the wave vector $k_\m$ since they were simply related
by raising the index using the spacetime metric, $p^\m = g^{\m\n}k_\n$. 
In the modified theory, however, there is an important distinction. 
The wave vector, defined as the gradient of the phase, is a covariant vector 
or 1-form, whereas the photon momentum/tangent vector to the rays is a true 
contravariant vector. The relation is non-trivial. 
In fact, given $k_\m$, we should define the corresponding
`momentum' as 
\begin{equation}
p^\m = {\cal G}^{\m\n}k_\n
\label{eq:bk}
\end{equation}
and the light rays as curves $x^\m(s)$ where ${dx^\m\over ds} = p^\m$.
This definition of momentum satisfies 
\begin{equation}
G_{\m\n} p^\m p^\n = {\cal G}^{\m\n}k_\m k_\n = 0
\label{eq:bl}
\end{equation}
where $G \equiv {\cal G}^{-1}$ defines a new {\it effective metric}
which determines the light cones mapped out by the geometric optics light rays.
(Indices are always raised or lowered using the true metric $g_{\m\n}$.)
The {\it ray velocity} $v_{\rm ray}$ corresponding to the momentum $p^\m$, 
which is the velocity with which the equal-phase surfaces advance, is given by
(defining components in an orthonormal frame)
\begin{equation}
v_{\rm ray} = {|\ul p|\over p^0} = {d |\ul x|\over dt}
\label{eq:bm}
\end{equation}
along the ray. This is in general different from the {\it phase velocity}
\begin{equation}
v_{\rm ph} = {\w\over|\ul k|}
\label{eq:bn}
\end{equation}
where the frequency $\w = k^0$.

This shows that photon propagation for low-energy QED in a background
electromagnetic or gravitational field (i.e.~curved spacetime) can be
characterised as a {\it bimetric} theory -- the physical light cones are
determined by the effective metric $G_{\m\n}$ while the geometric null
cones are fixed by the spacetime metric $g_{\m\n}$. 

This also clarifies a potentially confusing point. In the perturbative
situation we are considering here, where the effective Lagrangian 
$\LL(\FF,\GG) = -\FF + O(\a)$ and the effective metric $G_{\m\n}
= g_{\m\n} + O(\a)$, {\it subluminal} propagation with $v_{\rm ph} < 1$
corresponds to $k^2 < 0$, i.e.~the wave-vector is spacelike. However,
the analysis above makes it clear that in this case the photon momentum
is nevertheless timelike, i.e.~$p^2 > 0$ if $v_{\rm ph} < 1$ (and 
$v_{\rm ray} < 1$). For further discussion of this point and an 
illustration in the case of QED in a background FRW spacetime, 
see ref.\cite{Shore:2003jx}.

Now consider a general low-energy effective action $\LL(\FF,\GG)$. The
modified Maxwell equation is
\begin{equation}
\LL_\FF \partial_\m F^{\m\n} ~+~ \LL_\GG \partial_\m \tilde F^{\m\n} ~+~
\biggl(\LL_{\FF \FF} F^{\m\n} F_{\l\r} ~+~ \LL_{\GG \GG} \tilde F^{\m\n}
\tilde F_{\l\r} ~+~ \LL_{\FF \GG}\bigl(F^{\m\n} \tilde F_{\l\r} + 
\tilde F^{\m\n} F_{\l\r}\biggr) \partial_\m F^{\l\r} ~~=~~0
\label{eq:bo}
\end{equation}
while the Bianchi identity remains 
\begin{equation}
\partial_\m \tilde F^{\m\n} ~~=~~0
\label{eq:bp}
\end{equation}
We now immediately restrict to the case of interest here, viz.
$\LL(\FF,\GG) = -\FF + O(\a)$.  It would be trivial to extend the results of 
this section to this more general case, which includes the important example
of Born-Infeld theory.\footnote{The essential simplification we make is that
we can drop the factor $\LL_\FF$ in front of $\partial_\m F^{\m\n}$ since the
latter is already $O(\a)$ by virtue of the equations of motion, so we should
approximate $\LL_\FF \simeq 1$ when we work consistently to $O(\a)$.}
Applying geometric optics, we then find
\begin{eqnarray}
k^2 a^\n - k.a k^\n - \LL_{\FF \FF} k_\m F^{\m\n} F_{\l\r} k^\l a^\r 
- \LL_{\GG \GG}k_\m \tilde F^{\m\n} \tilde F_{\l\r} k^\l a^\r 
~~~~~~~~~~~~~~~~~~~~~~\nonumber\\
- \LL_{\FF \GG}\bigl(k_\m F^{\m\n} \tilde F_{\l\r} k^\l a^\r
+ k_\m \tilde F^{\m\n} F_{\l\r} k^\l a^\r\bigr) ~~=~~0
\label{eq:bq}
\end{eqnarray}
where the field strengths now refer to the background electromagnetic field.

The natural language to discuss photon propagation in background fields is the 
Newman-Penrose (NP), or null tetrad, formalism (see appendix A). At each point
in spacetime, we establish a tetrad of null vectors $\ell^\m, n^\n, m^\m, 
\bar m^\m$ such that $\ell.n = 1, m.\bar m = -1$ with all others zero.
We choose $\ell^\m$ to lie along the (unperturbed) direction of propagation,
i.e.~ $k^\m = \sqrt{2}\w \ell^\m + O(\a)$, where $\w$ is normalised to 
be the frequency. The polarisation vector is then 
\begin{equation}
a^\m ~~=~~ \a m^\m ~+~ \b \bar m^\m
\label{eq:br}
\end{equation}
where $m^\m (\bar m^\m)$ corresponds to left (right) circular polarisation,
i.e.~photon helicity $+1 (-1)$.

Substituting for the polarisation $a^\m$ in eq.(\ref{eq:bq}) and contracting
successively with $\bar m^\n$ and $m^\n$ then yields the equations:
\begin{equation}
k^2 \a ~+~ 2\w^2\Bigl(\LL_{\FF \FF}F_{\m\n}F_{\l\r} +
\LL_{\GG \GG}\tilde F_{\m\n} \tilde F_{\l\r} + 
\LL_{\FF \GG} (F_{\m\n} \tilde F_{\l\r} + \tilde F_{\m\n} F_{\l\r})\Bigr)
\ell^\m \ell^\l \bar m^\n (\a m^\r + \b \bar m^\r)~~=~~0
\label{eq:bs}
\end{equation}
and
\begin{equation}
k^2 \b ~+~ 2\w^2\Bigl(\LL_{\FF \FF}F_{\m\n}F_{\l\r} +
\LL_{\GG \GG}\tilde F_{\m\n} \tilde F_{\l\r} + 
\LL_{\FF \GG} (F_{\m\n} \tilde F_{\l\r} + \tilde F_{\m\n} F_{\l\r})\Bigr)
\ell^\m \ell^\l  m^\n (\a m^\r + \b \bar m^\r)~~=~~0
\label{eq:bt}
\end{equation}

To simplify further, we now insert the explicit form for the background fields
written in NP form. Here, the six independent components of the field
strength $F^{\m\n}$, i.e.~the ${\bf E}$ and ${\bf B}$ fields, are represented
by three complex scalars $\phi_0, \phi_1, \phi_2$ such that (see appendix A)
\begin{equation}
F^{\m\n} ~=~ -(\phi_1 + \phi_1^*) [\ell^\m n^\n] + 
(\phi_1 - \phi_1^*)[m^\m \bar m^\n] +\phi_2 [\ell^\m m^\n] 
+ \phi_2^* [\ell^\m \bar m^\n] - \phi_0^*[n^\m m^\n] - 
\phi_0 [n^\m \bar m^\n]  
\label{eq:bu}
\end{equation}
where the notation is $[a b] \equiv ab - ba$.
Then, after some calculation, we find that eqs.(\ref{eq:bs}), (\ref{eq:bt}) 
can be written in matrix form as
\begin{equation}
\left(\matrix{k^2 + 2\w^2 (\LL_{\FF \FF} + \LL_{\GG \GG})\phi_0 \phi_0^*
&2\w^2 (\LL_{\FF \FF} - \LL_{\GG \GG} - 2i \LL_{\FF \GG}) \phi_0^* \phi_0^*\cr
{}&{}\cr
2\w^2 (\LL_{\FF \FF} - \LL_{\GG \GG} + 2i \LL_{\FF \GG}) \phi_0 \phi_0
&k^2 + 2\w^2 (\LL_{\FF \FF} + \LL_{\GG \GG})\phi_0 \phi_0^*\cr}\right)~
\left(\matrix{\a\cr {}\cr \b\cr}\right) ~~=~~ 0
\label{eq:bv}
\end{equation}
The coefficients have a particularly natural interpretation in terms of the
variables
\begin{equation}
\XX ~=~ {1\over2}(\FF + i \GG)~~~~~~~~~~~~~~~~\bar\XX ~=~ 
{1\over2}(\FF -i\GG)  
\label{eq:bw}
\end{equation}
with eq.(\ref{eq:bv}) simplifying to
\begin{equation}
\left(\matrix{k^2 + 2\w^2 \LL_{\XX \bar\XX}\phi_0 \phi_0^*
&2\w^2 \LL_{\XX \XX} \phi_0^* \phi_0^*\cr
{}&{}\cr
2\w^2 \LL_{\bar\XX \bar\XX} \phi_0 \phi_0
&k^2 + 2\w^2 \LL_{\XX \bar\XX} \phi_0 \phi_0^*\cr}\right)~
\left(\matrix{\a\cr {}\cr \b\cr}\right) ~~=~~ 0
\label{eq:bx}  
\end{equation}

The dependence on the background field is especially illuminating.
The energy-momentum tensor is
\begin{equation}
T_{\m\n} ~~=~~ - F_{\m\l} F_\n{}^\l ~+~ {1\over4}g_{\m\n}F^2  
\label{eq:by}
\end{equation}
and in the NP basis, this implies
\begin{equation}
T_{\ell\ell} ~\equiv~T_{\m\n}\ell^\m \ell^\n ~=~ 2 \phi_0 \phi_0^*  
\label{eq:bz}
\end{equation}
This is of course the component of the energy-momentum tensor arising in 
the `{\it null energy condition}',
\begin{equation}
T_{\m\n} \ell^\m \ell^\n ~>~ 0  
\label{eq:baa}
\end{equation}
As we now see, it plays a vital role in the modified dispersion relations.

The dispersion relation follows directly from eq.(\ref{eq:bx}) as 
the requirement that the determinant of the matrix vanishes, or in other 
words, the allowed values of $k^2$ are the eigenvalues of the matrix
\begin{equation}
\left(\matrix{2\w^2 \LL_{\XX \bar\XX}\phi_0 \phi_0^*
&2\w^2 \LL_{\XX \XX} \phi_0^* \phi_0^*\cr
{}&{}\cr
2\w^2 \LL_{\bar\XX \bar\XX} \phi_0 \phi_0
&2\w^2 \LL_{\XX \bar\XX} \phi_0 \phi_0^*\cr}\right)  
\label{eq:bbb}
\end{equation}
This gives
\begin{equation}
k^2 ~~=~~ -2\w^2 \phi_0 \phi_0^* ~\Bigl(\LL_{\XX \bar \XX} \pm 
\sqrt{\LL_{\XX\XX} \LL_{\bar\XX \bar\XX}}~\Bigr)  
\label{eq:bcc}
\end{equation}
that is,
\begin{equation}
k^2 ~~=~~ -{1\over2} T_{\m\n} k^\m k^\n ~\Bigl(\LL_{\FF\FF} + \LL_{\GG\GG}
\pm \sqrt{(\LL_{\FF\FF}-\LL_{\GG\GG})^2 + 4 \LL_{\FF\GG}^2}~\Bigr)  
\label{eq:bdd}
\end{equation}

We see immediately what will later be established as a very general result,
viz.~that the modification in the light cone due to the background field
is proportional to the projection of the energy-momentum tensor onto the
null cone, i.e.~the component $T_{\m\n}k^\m k^\n$ which appears in the
null energy condition and whose sign is therefore fixed.  This dependence
turns out to be completely general and is the origin of the universal 
features of light propagation observed, for example, 
in ref.\cite{Latorre:1994cv}.\footnote{In fact, this analysis sharpens 
the conclusion of ref.\cite{Latorre:1994cv} by making clear that it is
the {\it null energy projection} $T_{\m\n}k^\m k^\n$ which controls the
dispersion relation rather than the energy density itself.
This observation clarifies a number of other results in the literature.
For example, in their paper \cite{Heinzl:2006xc} discussing the possibility 
of an experimental observation of electromagnetic birefringence,
Heinzl {\it et al.} \cite{Heinzl:2006xc} relate the low-frequency limit 
of the refractive index to `an energy density $Q^2$' where
$$
Q^2 ~=~ {\bf E}^2 + {\bf B}^2 - 2{\bf S}.{\hat {\bf k}} - 
({\bf E}.{\hat{\bf k}})^2 - ({\bf B}.{\hat{\bf k}})^2
$$
and ${\bf S}$ is the Poynting vector.
From the identities in the appendix, it is easy to check in our
conventions that $Q^2$ is indeed just $2~T_{\m\n}\ell^\m \ell^\n$.
(See also ref.\cite{Heinzl:2006pn}.)}

Notice also that as a direct consequence of its derivation from the
{\it low-energy} effective action, eq.(\ref{eq:bdd}) is homogeneous in $k^\m$
so there is no dispersion in this approximation. The result is of the
general form considered above in eq.(\ref{eq:bj}) and gives a phase velocity
\begin{equation}
v_{\rm ph}(0)~=~ 1 - {1\over2} T_{\m\n} \ell^\m \ell^\n ~\Bigl(\LL_{\FF\FF} 
+ \LL_{\GG\GG}\pm \sqrt{(\LL_{\FF\FF}-\LL_{\GG\GG})^2 
+ 4 \LL_{\FF\GG}^2}~\Bigr)    
\label{eq:bee}
\end{equation}

The corresponding polarisation vectors are the solutions for $\a,\b$ in 
eq.(\ref{eq:bv}), i.e.~the eigenvectors of (\ref{eq:bbb}). This gives
\begin{eqnarray}
&a_{+}^\m ~&=~ {1\over\sqrt2} 
\bigl( e^{i\theta} m^\m + e^{-i\theta}\bar m^\m\bigr) \nonumber\\
{}\nonumber\\
&a_{-}^\m ~&=~ -{i\over\sqrt2} 
\bigl( e^{i\theta} m^\m - e^{-i\theta}\bar m^\m\bigr)  
\label{eq:fff}
\end{eqnarray}
where the phase is defined from $\phi_0^* \sqrt{\LL_{\XX\XX}} = r e^{i\theta}$.
These combinations $a_{\pm}^\m$ of the left and right circular polarisations
represent orthogonal {\it linear} polarisations, with the direction fixed
by the angle $\theta$. The situation is therefore very similar to the case
of the polarisation eigenstates for propagation in a gravitational background
considered, e.g.~in ref.\cite{Shore:2003jx}.

Now, as explained above, in order to have {\it subluminal} propagation and
$v_{\rm ph} < 1$, we require $k^2 < 0$. Given the null energy condition
$T_{\m\n}k^\m k^\n > 0$, it follows from eqs.(\ref{eq:bcc}),(\ref{eq:bdd})
that this is ensured if 
\begin{equation}
{\rm det}~\left(\matrix{\LL_{\XX\bar\XX} & \LL_{\XX\XX} \cr
{}&{}\cr
\LL_{\bar\XX\bar\XX}& \LL_{\XX\bar\XX}\cr}\right) ~>~0,~~~~~~~~~~
\LL_{\XX\bar\XX} ~>~0
\label{eq:bgg}  
\end{equation}
or equivalently,
\begin{equation}
{\rm det}~\left(\matrix{\LL_{\FF\FF} & \LL_{\FF\GG} \cr
{}&{}\cr
\LL_{\FF\GG} & \LL_{\GG\GG} }\right) ~>~0, ~~~~~~~~~~
\LL_{\FF\FF} + \LL_{\GG\GG} ~>~ 0  
\label{eq:bhh}  
\end{equation}
These are the constraints placed on the coefficients of the leading
irrelevant operators in a general low-energy effective action by the
requirement of no superluminal propagation, at least (see section 3)
the requirement that $v_{\rm ph}(0) < 1$.

If we specialise to the Euler-Heisenberg action (\ref{eq:bb}), we can see
how this constraint is realised in low-energy effective QCD. We find
\begin{equation}
\LL_{\FF\FF} = {4\over45}{\a^2\over m^4} c_1,~~~~~~~~
\LL_{\GG\GG} = {4\over45}{\a^2\over m^4} c_2,~~~~~~~~
\LL_{\FF\GG} = 0
\label{eq:bii}
\end{equation}
and the positivity constraint reduces to simply $c_1 > 0, ~c_2 > 0$.
Of course this is satisfied by the QED values $c_1 = 4, ~c_2 = 7$.

The new light cone condition for the Euler-Heisenberg action is simply
\begin{equation}
k^2 ~+~ {4\over45}{\a^2\over m^4} T_{\m\n}k^\m k^\n ~[4_+, 7_-] ~~=~~ 0
\label{eq:bjj}
\end{equation}
corresponding to the linear polarisations $a_+^\m, a_-^\m$ respectively.
This reproduces the results found in our earlier paper \cite{Shore:1995fz}
where the dependence on the null energy component $T_{\m\n}k^\m k^\n$
was first recognised. It was observed there that the positivity of 
$c_1$ and $c_2$ also ensured the positivity of the trace of the 
energy-momentum tensor, but subsequent work \cite{Gies:1998sh}
has shown that this apparent link between causality and the conformal
anomaly is merely a low-order artifact. It also reproduces the classic
result for photon propagation in a purely magnetic background field
\cite{Tsai:1974fa,Tsai:1975iz}. 
In this case, it is straightforward to check that the
phase angle entering the formula for the polarisation eigenstates
is just $\tan \theta = B_1/B_2$ (where we take the direction of propagation
to be the $3$-axis), so that $a_-^\m$ lies parallel to ${\bf B}$
while $a_+^\m$ is orthogonal.

To summarise, working in low-energy electrodynamics, we have seen how
the light cone condition is modified by the component of the energy-momentum
tensor which is fixed by the null energy condition. The eigenstates of
definite phase velocity are linear polarisations with direction fixed
by the background field. Ensuring $v_{\rm ph} < 1$ imposes positivity
constraints on the coefficients of the leading irrelevant operators in
the effective action. The modified light cone condition is homogeneous
and quadratic in $k$ and there is no dispersion.

However, the absence of dispersion is merely an artifact of the low-energy
approximation. In general, there {\it is} non-trivial dispersion and the
light cone condition (\ref{eq:bdd}) only fixes the {\it low-frequency limit}
$v_{\rm ph}(0)$ of the phase velocity. In the following section, we discuss
more carefully what causality actually requires and find that, far from
imposing any constraint on the low-frequency behaviour of the phase velocity,
it imposes a constraint on the {\it high-frequency limit}, 
$v_{\rm ph}(\infty) < 1$. In turn, this is controlled {\it not} by the 
low-energy effective action but by the UV limit of the quantum field theory.

\section{Speeds of light, causality and dispersion relations}

In order to understand whether causality does indeed impose constraints on 
the parameters of a low-energy effective action, we first need to 
understand more precisely exactly what we mean by the `speed of light'.

By definition, a low-energy effective action is an expansion in powers
of derivatives acting on the fields. For example, the Euler-Heisenberg
effective action (\ref{eq:ba}) contains only terms of $O(F^4/m^4)$ leading, 
as we have seen, to a homogeneous dispersion relation for photon propagation
with each term of $O(k^2)$.
However, this neglects higher derivative terms beginning with 
$O(D^2 F^4/m^6)$, which modify the dispersion relation with terms 
suppressed by $O(k^2/m^2)$. This is why the `speed of light' derived in 
section 2 from the low-energy effective action is just the low-frequency
limit of the phase velocity, i.e.~$v_{\rm ph}(\w)|_{\w\rta 0}$.

In fact, there are many `speeds of light', of which the phase velocity is 
just one. A comprehensive account is given in the classic text by
Brillouin \cite{Brillouin}, which considers in detail the propagastion
of a sharp-fronted wave pulse in a medium with a single absorption
band (see below). The essential ideas are explained in our
review \cite{Shore:2003jx}. As well as the phase velocity 
$v_{\rm ph}(\w) = {\w\over |\underline k|}$, we need to consider the
group velocity $v_{\rm gp}(\w) = {d\w\over d|\underline k|}$ and the
wavefront velocity, i.e.~the velocity at which the boundary between
the regions of zero and non-zero disturbance of a wave
pulse propagates. This definition of wavefront is identified with 
the characteristics of the partial differential equation governing the
wave propagation and it is the wavefront velocity $v_{\rm wf}$ is the 
speed of light which is relevant for causality.

Other definitions of the `speed of light', each of which is useful in
a particular context, can be found in ref.\cite{Shore:2003jx}, e.g.
signal velocity, energy-transfer velocity, etc. In addition, as we
have seen in section 2, the ray velocity also plays a conceptually 
important role.

We now show how the wavefront velocity is related to the phase
velocity. The following proof follows a paper by Leontovich
\cite{Leontovich}, as summarised in ref.\cite{Shore:2003jx}. We show that
for a very general set of PDEs the wavefront velocity associated with 
the characteristics is identical to the $k\rta\infty$ limit of the 
phase velocity, i.e.
\begin{equation}
v_{\rm wf} = \lim_{k\rta \infty}{\w\over|\ul k|} = 
\lim_{\w\rta \infty}v_{\rm ph}(\w)
\label{eq:ca}
\end{equation}

The first step is to recognise that any second order PDE can be written as
a system of first order PDEs by considering the first derivatives of
the field as independent variables. Thus, if we consider a general second 
order wave equation for a field $u(t,x)$, in just one space dimension for
simplicity, the system of PDEs we need to solve is
\begin{equation}
a_{ij} {\partial\phi_j\over\partial t} + b_{ij} 
{\partial\phi_j\over\partial x}
+ c_{ij} \phi_j ~=~0
\label{eq:cb}
\end{equation}
where $\phi_i = \{u, {\partial u\over\partial t}, 
{\partial u\over\partial x}\}$.

Making the `geometric optics' ansatz
\begin{equation}
\phi_i ~=~ \varphi_i \exp i(wt-kx)
\label{eq:cc}
\end{equation}
where the frequency-dependent phase velocity is $v_{\rm ph}(k) = \w(k)/k$, 
and substituting into eq.(\ref{eq:cb}) we find
\begin{equation}
\Bigl(i\w a_{ij} - ik b_{ij} + c_{ij} \Bigr) \varphi_j ~=~ 0
\label{eq:cd}
\end{equation}
The condition for a solution,
\begin{equation}
{\rm det}\Bigl[a_{ij} v_{\rm ph}(k) - b_{ij} -{i\over k} c_{ij} \Bigr] ~=~ 0
\label{eq:ce}
\end{equation}
then determines the phase velocity.

On the other hand, we need to find the characteristics of eq.(\ref{eq:cb}),
i.e.~curves ${\cal C}$ on which Cauchy's theorem breaks down and the 
evolution is not uniquely determined by the initial data on ${\cal C}$. 
The derivatives of the field may be discontinuous across the
characteristics and these curves are associated with the wavefronts
for the propagation of a sharp-fronted pulse. The corresponding light rays
are the `bicharacteristics'.  

We therefore consider a characteristic curve ${\cal C}$ in the $(t,x)$
plane separating regions where $\phi_i =0$ (ahead of the wavefront)
from $\phi_i \ne 0$ (behind the wavefront). At a fixed point
$(t_0,x_0)$ on ${\cal C}$, the absolute derivative of 
$\phi_i$ along the curve, parametrised as $x(t)$, is just
\begin{equation}
{d\phi_i\over dt} ~=~ {\partial \phi_i\over\partial t}\Big|_0
+ {\partial \phi_i\over\partial x}\Big|_0 {dx\over dt}
\label{eq:cf}
\end{equation}
where $dx/dt = v_{\rm wf}$ gives the wavefront velocity. 
Using this to eliminate ${\partial\phi_i\over\partial t}$ from the PDE 
eq.(\ref{eq:cb}) at $(t_0,x_0)$, we find
\begin{equation}
\Bigl( - a_{ij} {dx\over dt} + b_{ij}\Bigr) 
{\partial \phi_j\over\partial x}\Big|_0
+ a_{ij} {d\phi_j^{(0)}\over dt} + c_{ij}\phi_j^{(0)} ~=~ 0
\label{eq:cg}
\end{equation}
Now since ${\cal C}$ is a wavefront, on one side of which $\phi_i$ vanishes
identically, the second two terms above must be zero.
The condition for the remaining equation to have a solution is simply
\begin{equation}
{\rm det}\Bigl[ a_{ij} v_{\rm wf} - b_{ij} \Bigr] ~=~ 0
\label{eq:ch}
\end{equation}
which determines the wavefront velocity $v_{\rm wf}$.
The proof is now evident. Comparing eqs.(\ref{eq:ce}) and (\ref{eq:ch}), we
clearly identify
\begin{equation}
v_{\rm wf} ~=~ v_{\rm ph}|_{k\rta \infty}
\label{eq:ci}
\end{equation}

The conclusion is that the wavefront velocity is in fact the 
{\it high-frequency} limit of the phase velocity. That is, the 
`speed of light' which is relevant for causality is {\it not} the 
low-frequency phase velocity $v_{\rm ph}(0)$ but its high-frequency 
limit $v_{\rm ph}(\infty)$.

In flat spacetime, it is therefore natural, and indeed true, that 
causality requires $v_{\rm ph}(\infty) < 1$. In curved spacetime, 
however, the requirement of causality is much more subtle. Indeed, as
explained in ref.\cite{Shore:2003jx} 
(see also ref.\cite{Visser:1998ua,Liberati:2001sd}),
it is in principle possible to maintain a causal theory even with
$v_{\rm ph}(\infty) > 1$ provided that the background spacetime exhibits
a modified version of `stable causality'. This possible loophole is most 
easily illustrated for the case considered in section 2 where the 
apparently superluminal wave equation is homogeneous and quadratic, 
although the physical idea remains relevant for general propagation 
equations. In this case, the dispersion relation may be written as 
$\GG^{\m\n}k_\m k_\n = 0$. As explained in section 2, the effective
light cone is then determined by an `effective metric' $G_{\m\n}$, 
where $G = \GG^{-1}$, which depends on the background fields and 
which we think of as being perturbatively close to the spacetime 
metric $g_{\m\n}$. This generalises the standard general relativity 
assumption that light follows null geodesics, in which case the 
dispersion relation is just $g^{\m\n}k_\m k_\n = 0$.

The concept of {\it stable causality} is summarised in the following 
definition and theorem \cite{HawkingEllis}:

\noindent $\bullet$ A spacetime manifold $({\cal M},g_{\m\n})$ is 
{\it stably causal} if the metric $g_{\m\n}$ has an open neighbourhood 
such that ${\cal M}$ has no closed timelike or null curves with respect 
to any metric belonging to that neighbourhood.

\noindent $\bullet$ Stable causality holds everywhere on ${\cal M}$ if and 
only if there is a globally defined function $f$ whose gradient $D_\m f$ 
is everywhere non-zero and timelike with respect to $g_{\m\n}$.   

According to this theorem, the absence of causality violation in the form 
of closed timelike or lightlike curves is assured if we can find a globally 
defined function $f$ whose gradient is timelike {\it with respect to the 
effective metric} $G_{\m\n}$ for light propagation. $f$ then acts as a 
global time coordinate. That this can occur is demonstrated for the
example of `superluminal' light propagation in a Friedmann-Robertson-Walker
spacetime in ref.\cite{Shore:2003jx}.

Setting this curved-spacetime subtlety aside for the moment, our conclusion 
here is that the restriction imposed by causality on the propagation of 
light is {\it not} concerned with the low-frequency, IR limit $v_{\rm ph}(0)$ 
determined by the low-energy effective action. Rather, causality imposes a 
bound on the {\it high-frequency} limit, $v_{\rm ph}(\infty) < 1$. This is 
determined by the UV nature of the quantum field theory.

At first sight, therefore, it appears that causality imposes no constraint
on the structure of low-energy effective theories. However, there is one
further twist. The (complex) refractive index $n(\w)$ satisfies a 
Kramers-Kronig (KK) dispersion relation\footnote{In what follows, 
we always refer to this as the `KK dispersion relation' to avoid any 
possible confusion with our use so far of `dispersion relation' 
to describe the dependence of the frequency 
$\w(k)$ on wave-number $k$ determined by the light-cone condition.
We also drop the convention $c=1$ in this section.}
\begin{equation}
n(\infty) ~~=~~n(0) ~-~ {2\over\pi} \int_0^\infty {d\w\over\w}~ \Ima n(\w) 
\label{eq:cj}
\end{equation}
In a conventional dispersion relation, the imaginary part of the 
refractive index, i.e.~the absorption coefficient, is positive, 
$\Ima n(\w) > 0$. This implies 
that $n(\infty) < n(0)$ and so, recalling $v_{\rm ph}(\w) = 
1/\Rea n(\w)$, that $v_{\rm ph}(\infty) > v_{\rm ph}(0)$. 
If so, the phase velocity deduced from the low-energy effective action
would be a {\it lower bound} on the wavefront velocity $v_{\rm wf}
= v_{\rm ph}(\infty)$ and so a superluminal $v_{\rm ph}(0)$ would 
indeed represent a violation of causality. 

Similar KK dispersion relations arise in quantum field theory applied
to correlation functions and scattering amplitudes $\MM(s,t)$.
The axiomatic inputs into their derivation are the standard ones
of local quantum field theory, notably micro-causality, viz.~the 
vanishing of commutators of field operators evaluated at spacelike
separated points. In conventional QFT, the optical theorem relates 
the imaginary part $\Ima \MM(s,0)$ of a forward scattering amplitude 
via unitarity to the total cross section, so positivity is assured.

It therefore appears that positivity of $\Ima n(\w)$, and consequently
$v_{\rm ph}(\infty) > v_{\rm ph}(0)$, is guaranteed in any situation
of interest. This would ensure the validity of causality bounds derived
from the low-energy effective action. However, this conclusion may
be too fast. For example, consider an optical medium with a single
absorption band. Its refractive index may be modelled as \cite{Brillouin}
\begin{equation}
n^2(\w) ~~=~~ 1 ~-~ {M^2\over {\w^2 - \w_0^2 + i\d\w}}
\label{eq:ck}
\end{equation}
where $M$ determines the absorption strength, $\w_0$ the characteristic
frequency of the medium and $\d$ the width of the absorption band. 
The real and imaginary parts of $n(\w)$ are sketched in Fig.~1.

\FIGURE
{\epsfxsize=7cm\epsfbox{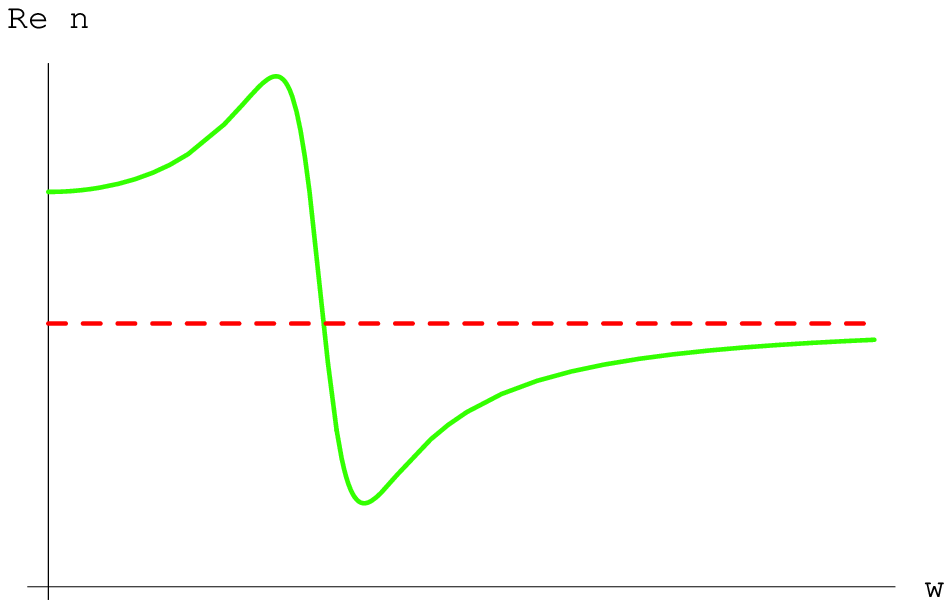}\hskip1cm
\epsfxsize=7cm\epsfbox{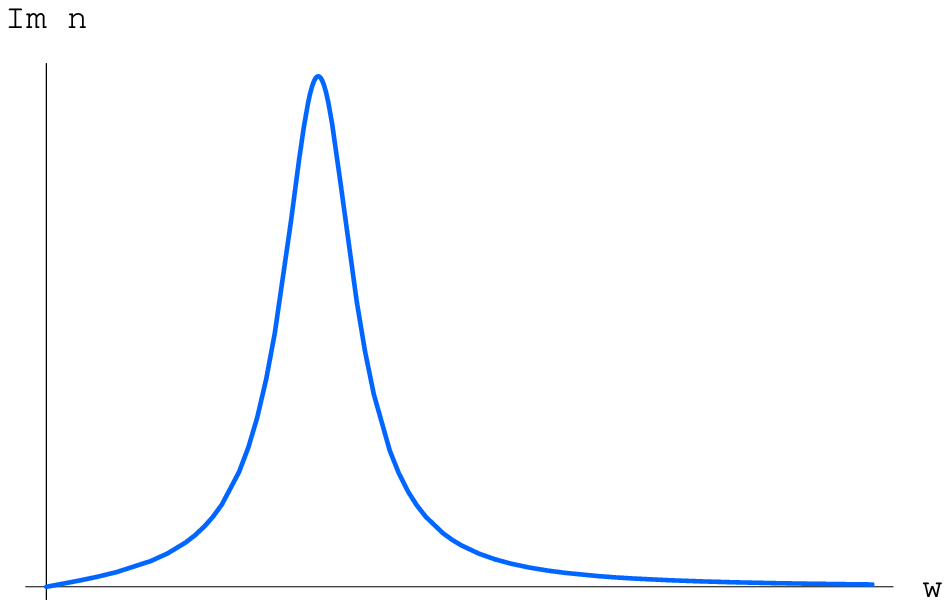}
\caption{The left hand figure shows the real part of the refractive index
$\Rea n(\omega)$ for the single absorption band model (\ref{eq:ck}). The 
dashed line is $n(\omega) = 1$.
The right hand plot is the corresponding $\Ima n(\omega)$.}
}

\FIGURE
{\epsfxsize=7cm\epsfbox{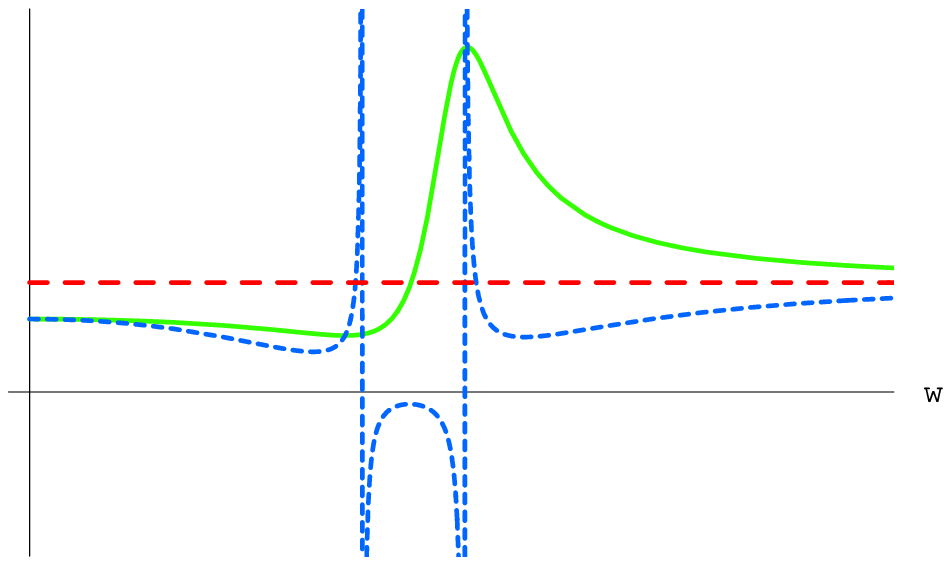}\hskip1cm
\epsfxsize=7cm\epsfbox{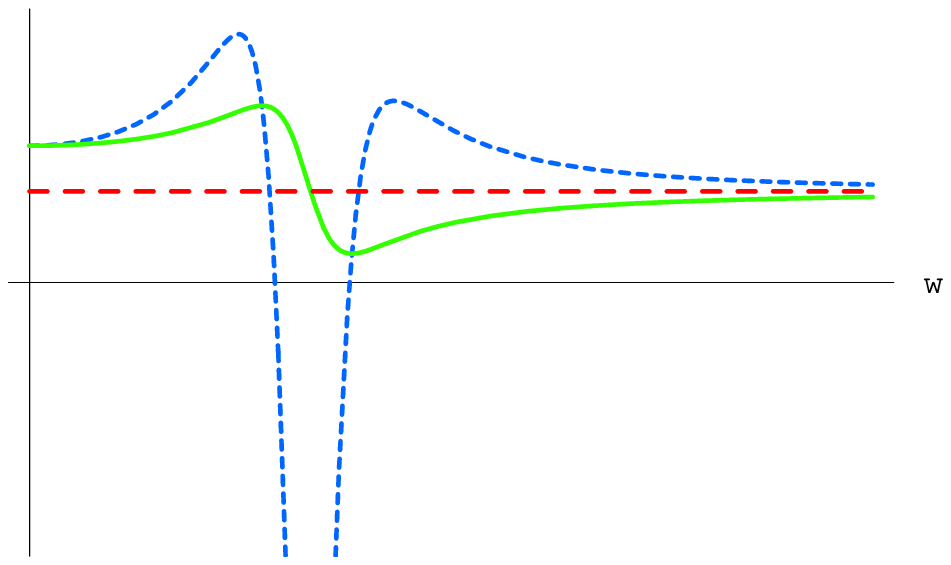}
\caption{The left hand figure shows the frequency dependence of the
phase velocity $v_{\rm ph}(\omega)$ (solid green line) and the group
velocity $v_{\rm gp}(\omega)$ (dashed blue line) for the single 
absorption band model. The constant wavefront velocity $v_{\rm wf}=1$ 
is indicated by the long-dashed red line. The right hand 
figure shows the corresponding indices, $n_{\rm ph}= \Rea n(\omega) 
= 1/v_{\rm ph}$ and $n_{\rm gp}(\omega) = 1/v_{\rm gp}$.}
}

This is the standard situation. The imaginary part $\Ima n(\w)$ is positive,
while $\Rea n(\w)$ initially rises (`normal dispersion') then falls rapidly
through the absorption band (`anomalous dispersion') passing through
$\Rea n = 1$ at the characteristic frequency $\w_0$ before asymptotically 
approaching $n(\infty) = 1$ from below. In terms of the phase velocity,
this gives $v_{\rm ph}(0) < 1$ and $v_{\rm ph}(\infty) = v_{\rm wf} = 1$,
with an interesting behaviour in the vicinity of the characteristic frequency
$\w_0$. The bound $v_{\rm ph}(\infty) > v_{\rm ph}(0)$ clearly holds.
In the following section, we derive the explicit behaviour of 
$v_{\rm ph}(\w)$ for QED in a background electromagnetic field and show 
how the above features are realised in a QFT context.

It will be clear, however, that all this would be reversed if we were
in a situation where $\Ima n(\w) < 0$. In fact, this can easily be realised
in, for example, atomic physics systems such as we describe in section 6.
It corresponds to the system exhibiting {\it gain} rather than absorptive 
scattering. The big question is whether such a phenomenon can arise
in QFT. In section 5, we show that for QED in a gravitational background
we can indeed find a superluminal low-frequency phase velocity
$v_{\rm ph}(0) > 1$. This poses a clear dilemma in what is after all a
very conventional QFT, albeit involving gravity: either we really find
$v_{\rm ph}(\infty) > v_{\rm ph}(0) > 1$ in which case we would be forced
to reassess fundamental issues regarding causality, or $\Ima n(\w) < 0$,
i.e.~propagation and scattering in a gravitational field can exhibit the 
characteristics of gain as well as dispersion.

All this will be explored in detail in future sections. It is immediately
clear, however, that we need to be extremely careful before drawing the
conclusion that apparently causality-violating values of couplings in
low-energy effective actions necessarily imply the absence of a well-defined
UV completion within the usual axioms of QFT, possibly including gravity.

\section{The QED refractive index}

In this section, we find an explicit representation of the dispersion 
relation for photon propagation in QED in a constant background 
electromagnetic field. This shows how the frequency dependence of the
refractive index described above is actually realised in a well-understood
quantum field theory. 

The field-theoretic calculations required to evaluate an effective action
valid for all momenta in a general background field are still beyond current 
techniques, so here we restrict the discussion to constant background fields.
Also, since we are concerned only with analysing photon propagation, it
is sufficient to evaluate the vacuum polarisation tensor (the second 
functional derivative of the effective action) in the given background.
The state of the art in such calculations are those of Schubert and Gies
\cite{Schubert:2000yt,Schubert:2000kf,Gies:2001zm} 
using the worldline path integral approach to QFT. This generalises the
earlier classic work on QED in background fields using the more
conventional Schwinger proper time or heat kernel methods (see for example
\cite{Schwinger:1951nm,Tsai:1974fa,Tsai:1975iz}). 
Here, we start from the result obtained
in ref.\cite{Schubert:2000yt} for the vacuum polarisation tensor in a constant
background electromagnetic field and translate it into the NP formalism
before considering in detail its consequences for photon propagation,
including the realisation of the KK dispersion relation.

Our starting point is therefore the formula
(eq.(4.11) of ref.\cite{Schubert:2000yt}) 
for the one-loop QED vacuum polarisation:
\begin{eqnarray}
\Pi_{\m\n}(k) ~~=~~ {\a\over4\pi} \int_0^\infty {ds\over s} e^{-ism^2}
\int_{-1}^1 dv \biggl[ {z_+ z_-\over \tanh z_+ \tanh z_-}~~ 
 \exp\Bigl(-i{s\over2} \sum_{\a = +,-} C_\a~ k.Z_\a^2.k\Bigr)~\times
~~~~~~~~~~~~~\nonumber \\
\sum_{\a,\b=+,-} \Bigl(S_{\a\b} \Bigl[(Z_\a^2)_{\m\n} k.Z_\b^2 .k
- (Z_\a^2 k)_\m (Z_\b^2 k)_\n \Bigr] ~+~ A_{\a\b}~ (Z_\a k)_\m (Z_\b k)_\n
 ~-~ (g_{\m\n} k^2 - k_\m k_\n)(v^2 -1)\Bigr) \biggr]\nonumber \\
{}\nonumber\\
\label{eq:da}
\end{eqnarray}
where
\begin{equation}
z_+ = iesa ~~~~~~~~z_- = -esb ~~~~~~~~ Z_+ = {aF - b\tilde F \over a^2 + b^2}
~~~~~~~~Z_- = -i{bF + a\tilde F \over a^2 + b^2}
\label{eq:db}
\end{equation}
with
\begin{equation}
a = \sqrt{\sqrt{\FF^2 + \GG^2} + \FF} ~~~~~~~~
b = \sqrt{\sqrt{\FF^2 + \GG^2} - \FF}
\label{eq:dc}
\end{equation}
The explicit expressions for the dynamical coefficients $C_\a, S_{\a\b},
A_{\a\b}$ are given later.

The dispersion relation for photon propagation is
\begin{equation}
k^2 a_\n ~-~ k.a k_\n ~-~ \Pi_{\m\n}a^\m ~~=~~0  
\label{eq:dd}
\end{equation}
Clearly, given the vacuum polarisation in the form (\ref{eq:da}), it is
far from transparent what its implications are for the physics of photon 
propagation. However, simplifications and physical meaning rapidly become
apparent as we translate into the null tetrad, NP form. In fact, it may
well be that the most efficient method for calculating vacuum polarisation
in background fields would be to adopt the NP formalism from the outset.
Indeed, this may offer the only realistic hope of reproducing similar
vacuum polarisation calculations for background gravitational fields
(see section 5).

\subsection{General structure of the vacuum polarisation and 
dispersion relations}

As before, we take $k^\m = \sqrt{2} \w \ell^\m + O(\a)$ and polarisation
vectors $a^\m = \a m^\m + \b \bar m^\m$ and look for the eigenvectors
and eigenvalues of the dispersion relation. Contracting with $m_\n, 
\bar m_\n$ in eq.(\ref{eq:dd}), we have
\begin{equation}
\left(\matrix{k^2 + \Pi_{m \bar m} & \Pi_{\bar m \bar m} \cr
\Pi_{m m}& k^2 + \Pi_{\bar m m} \cr}\right)
 \left(\matrix{\a \cr \b \cr}\right) ~=~0
\label{eq:de}
\end{equation}
where $\Pi_{m \bar m} = \Pi_{\m\n} m^\m \bar m^\n$ etc.

The first step is to translate the various `kinematic' terms in the vacuum
polarisation tensor into NP form. The necessary intermediate formulae are
collected in the appendix. For the required contractions, we find
\begin{eqnarray}
m.Z_{\pm}.\ell ~=~ {a-ib\over a^2 + b^2} ~\phi_0 ~~~~~~~~~~~~~~~
\bar m .Z_{\pm}.\ell ~=~ \mp {a+ib\over a^2 + b^2} ~\phi_0^* ~~~~~~~~~~~~
~~~~~~~~~~~~~~~~~~~~~\nonumber \\
{}\nonumber \\
\ell.Z_{\pm}^2 .\ell ~=~ \pm {1\over a^2 + b^2}~ 2\phi_0 \phi_0^* ~~~~~~~
m.Z_{\pm}^2 .\ell ~=~ \mp {1\over a^2 + b^2}~ 2\phi_0 \phi_1^* ~~~~~~~
\bar m .Z_{\pm}^2 .\ell ~=~ \mp {1\over a^2 + b^2} ~2\phi_1 \phi_0^*
\nonumber \\
{}\nonumber\\
m.Z_{\pm}^2 .m ~=~ \bar m .Z_{\pm}^2 .\bar m ~=~ 
\pm {1\over a^2 + b^2} ~2\phi_0 \phi_2^* ~~~~~~~~~~~~
m.Z_{\pm}^2 .\bar m ~=~ \bar m Z_{\pm}^2 .m ~=~ 
{1\over2} \pm {1\over a^2 + b^2}~2\phi_1 \phi_1^* 
\nonumber \\
{}
\label{eq:df}
\end{eqnarray}
where we have also used the results:
\begin{eqnarray}
a^2 + b^2 = 2\sqrt{\FF^2 + \GG^2} = 4\sqrt{\XX \bar\XX} 
~~~~~~~~~~~~~~~~~~~~~~~~~~~~~~~~~~~\nonumber \\
a^2 - b^2 = 2\FF = 4~ \Rea (\phi_0 \phi_2 - \phi_1^2) 
~~~~~~~~~~~~~~~~
2ab = 2\GG = 4~ \Ima (\phi_0 \phi_2 - \phi_1^2) 
\label{eq:dg}
\end{eqnarray}
This immediately begins to show the simplifications which arise by using
the NP components of the electromagnetic field strengths.
Several remarkable cancellations now occur. To illustrate with just 
two examples, we find:
\begin{eqnarray}
S_{++} \Bigl[(Z_+^2)_{mm} ~\ell.Z_+^2 .\ell ~-~ 
(Z_+^2 \ell)_m (Z_+^2 \ell)_m \Bigr] \nonumber \\
=~4\phi_0^2 ~{1\over (a^2 + b^2)^2} ~ \Bigl(\phi_0^* \phi_2^* 
- (\phi_1^*)^2\Bigr)~
S_{++} ~~~~~~~~~~\nonumber \\
=~{1\over4\XX} ~\phi_0^2 ~S_{++}~~~~~~~~~~~~~~~~~~~~~~~~~~~~~~~~~~~~~~~~~
\label{eq:dh}
\end{eqnarray}
while
\begin{eqnarray}
S_{++} \Bigl[(Z_+^2)_{\bar m m} ~\ell.Z_+^2 .\ell ~-~ 
(Z_+^2 \ell)_{\bar m} (Z_+^2 \ell)_m \Bigr] 
~~~~~~~~~~~~~~~~~~~~~~~~~~~~~~~~~~~~~~~~~~~\nonumber \\
=~\Bigl[ \Bigl({1\over2}~+~ {1\over a^2 + b^2}~2\phi_1 \phi_1^*\Bigr)
~{1\over a^2 + b^2}~2\phi_0 \phi_0^* ~-~ 
{1\over a^2 + b^2}~2\phi_1 \phi_0^* ~{1\over a^2 + b^2}~2\phi_0 \phi_1^*
\Bigr]~S_{++} \nonumber \\
=~ {1\over 4 \sqrt{\XX \bar \XX}} ~\phi_0 \phi_0^* ~S_{++}~~~~~~~~~~
~~~~~~~~~~~~~~~~~~~~~~~~~~~~~~~~~~~~~~~~~~~~~~~~~~~~~~~~~~~~~~~~~~~~~
\label{eq:di}
\end{eqnarray}

Collecting terms and substituting back into eq.(\ref{eq:da}), we therefore
find the following simplified form for the relevant components of the
vacuum polarisation tensor in the NP basis:
\begin{equation}
\Pi_{AB}(k) ~~=~~ {\a\over4\pi} \int_0^\infty {ds\over s} e^{-ism^2}
\int_{-1}^1 dv ~ {z_+ z_-\over \tanh z_+ \tanh z_-}~~ 
 \exp\Bigl(-is\w^2 ~\phi_0 \phi_0^*~{(C_+ - C_-)\over2\sqrt{\XX\bar \XX}}~
\Bigr)~~ \tilde \Pi_{AB} 
\label{eq:dj}
\end{equation}
where
\begin{eqnarray}
\tilde \Pi_{\bar m m} ~=~ \w^2 \phi_0 \phi_0^* ~
{1\over 2\sqrt{\XX \bar\XX}}~
\Bigl(S_{++} - S_{+-} + S_{-+} - S_{--} + 
A_{++} + A_{+-} - A_{-+} - A_{--} \Bigr)~
\nonumber \\
\tilde \Pi_{m \bar m} ~=~ \w^2 \phi_0 \phi_0^* ~
{1\over 2\sqrt{\XX \bar\XX}}~
\Bigl(S_{++} - S_{+-} + S_{-+} - S_{--} + 
A_{++} - A_{+-} + A_{-+} - A_{--} \Bigr)~
\nonumber \\
\tilde \Pi_{m m} ~=~ \w^2 \phi_0 \phi_0 ~
{1\over 2\XX}~
\Bigl(S_{++} - S_{+-} - S_{-+} + S_{--} + 
A_{++} + A_{+-} + A_{-+} + A_{--} \Bigr) ~~~~~~
\nonumber \\
\tilde \Pi_{\bar m \bar m} ~=~ \w^2 \phi_0^* \phi_0^* ~
{1\over 2\bar\XX} ~
\Bigl(S_{++} - S_{+-} - S_{-+} + S_{--} + 
A_{++} - A_{+-} - A_{-+} + A_{--} \Bigr) ~~~~~~
\label{eq:dk}
\end{eqnarray}

Notice that up to this point we have assumed nothing about the explicit
form of the coefficients $C_\a, S_{\a\b}$ and $A_{\a\b}$.  Self-consistency
is guaranteed, however, by certain explicit properties, e.g.~both $A_{\a\b}$
and $S_{\a\b}$ are symmetric on $\a,\b$ while all $S_{\a\b}$ are real,
$A_{++}, A_{--}$ are real and $A_{+-}, A_{-+}$ are pure imaginary.
Imposing these properties, we therefore find in the $m, \bar m$ sector:
\begin{equation}
\left(\matrix{\tilde \Pi_{m \bar m} & \tilde \Pi_{\bar m \bar m} \cr
\tilde \Pi_{m } &\tilde \Pi_{\bar m m}}\right) ~~=~~
{1\over 2}\w^2~
\left(\matrix{\AA ~{1\over\sqrt{\XX\bar\XX}}~\phi_0 \phi_0^* 
& \BB^*~{1\over \bar\XX}~\phi_0^* \phi_0^* \cr
\BB ~{1\over \XX}~\phi_0 \phi_0 
& \AA~{1\over\sqrt{\XX\bar\XX}}~\phi_0 \phi_0^*}\right)
\label{eq:dl}
\end{equation}
with
\begin{equation}
\AA~=~ S_{++} - S_{--} + A_{++} - A_{--}
\label{eq:dm}
\end{equation} 
and
\begin{equation}
\BB ~=~ S_{++} - 2S_{+-} + S_{--} + A_{++} + 2A_{+-} + A_{--}
\label{eq:dn}
\end{equation}
This matrix has eigenvalues
\begin{equation}
{1\over2}\w^2~\phi_0 \phi_0^*~{1\over \sqrt{\XX\bar\XX}} ~\Bigl(\AA \pm
\sqrt{\BB \BB^*} \Bigr)
\label{eq:do}
\end{equation}

Finally, putting all this together, we find the dispersion relation
for photon propagation in a constant background electromagnetic field,
as determined by the vacuum polarisation tensor (\ref{eq:da}), is:
\begin{eqnarray}
k^2 ~+~ {\a\over4\pi} \int_0^\infty {ds\over s} e^{-ism^2}
\int_{-1}^1 dv ~ {z_+ z_-\over \tanh z_+ \tanh z_-}~~ 
 \exp\Bigl(-is~{1\over8}T_{\m\n}k^\m k^\n~{1\over\sqrt{\XX\bar\XX}}~
(C_+ - C_-) \Bigr)
\nonumber \\ 
{}\nonumber\\
\times~\Bigl[{1\over8}~T_{\m\n}k^\m k^\n~{1\over\sqrt{\XX\bar\XX}}~  
\Bigl(\AA \pm \sqrt{\BB \BB^*} \Bigr) \Bigr]~~=~~ 0~~~~~~~~
\label{eq:dp}
\end{eqnarray}
corresponding (see section 2) to linear polarisation vectors
\begin{eqnarray}
&a^\m_+ ~&=~ {1\over\sqrt2} \Bigl( e^{i\theta} m ~+~ 
e^{-i\theta}\bar m\Bigr) \nonumber\\
&a^\m_- ~&=~ -{i\over\sqrt2} \Bigl( e^{i\theta} m ~-~ 
e^{-i\theta}\bar m\Bigr)
\label{eq:dq}
\end{eqnarray}
with the phase defined by $\phi_0^* \sqrt{{\BB^*\over\bar\XX}} 
= r e^{i\theta}$.

The most remarkable feature of this result is the appearance of the
{\it null energy} combination $T_{\m\n}k^\m k^\n = 4\w^2 \phi_0 \phi_0^*$,
both as an overall factor just as in the low-frequency limit
and also in the exponent which, as we shall see, controls the 
high-frequency limit. This is an exceptionally clear demonstration
of the direct relation, valid for {\it all} frequencies or momenta,
between the null energy condition and the presence or absence of
superluminal propagation of light.

\subsection{Dynamics and the frequency dependence of the refractive index}

Having established that the relation between photon propagation and the
null energy condition remains valid for all frequencies, the next step is
to investigate the detailed dynamics of the dispersion relation.
This is encoded in the coefficients $C_\a, S_{\a\b}$ and $A_{\a\b}$,
which are functions only of the two Lorentz invariant combinations of
the background electromagnetic field strengths, $\FF$ and $\GG$.

The required formulae can be extracted from the results derived in
ref.\cite{Schubert:2000yt}. We find
\begin{eqnarray}
C_\a ~=~ -{1\over z_\a} \biggl({\cosh z_\a v \over \sinh z_\a} - \coth z_\a
\biggr)
~~~~~~~~~~~~~~~~~~~~~~~~~~~~~~~~~~~~~~~~~~~~\nonumber\\
{}\nonumber\\
S_{\a\b} ~=~-{\cosh z_\a v \over \cosh z_\a}{\cosh z_\b v \over \cosh z_\b}  
+ {\sinh z_\a v \over \sinh z_\a}{\sinh z_\b v \over \sinh z_\b}
~~~~~~~~~~~~~~~~~~~~~~~~~~~~\nonumber\\
{}\nonumber\\
A_{\a\b} ~=~ -\biggl({\cosh z_\a v \over \sinh z_\a} - \coth z_\a
+ \tanh z_\a \biggr) \biggl( \a \leftrightarrow \b \biggr) ~+~
{\sinh z_\a v \over \cosh z_\a} {\sinh z_\b v \over \cosh z_\b} 
\label{eq:dr}
\end{eqnarray}
Expanding these in a weak field expansion, we find \footnote{
The dictionary between our notation and
that of ref.\cite{Schubert:2000yt} is:
\begin{eqnarray*}
C_\a = -{1\over z_\a} (A_{B12}^\a - A_{B11}^\a) 
~~~~~~~~~~~~~~~~~~~~~~~~~~~~~~~~~~~~~~~~~~~~~~~~~~\nonumber\\
S_{\a\b} = S_{B12}^\a S_{B12}^\b - S_{F12}^\a S_{F12}^\b 
~~~~~~~~~~~~~~~~~~~~~~~~~~~~~~~~~~~~~~~~~~~~~~~\nonumber\\
A_{\a\b} = -\Bigl[(A_{B12}^\a - A_{B11}^\a + A_{F11}^\a)
(A_{B12}^\b - A_{B22}^\b + A_{F22}^\b)\Bigr] + A_{F12}^\a A_{F12}^\b 
\nonumber\\
\end{eqnarray*}
where the worldline formalism coefficient functions are:
\begin{eqnarray*}
S_{B12}^\a = {\sinh z_\a v \over \sinh z_\a} ~~~~~~~~~~~~~~~~~~~~~~~~~~~~
S_{F12}^\a = G_{F12} {\cosh z_\a v \over \cosh z_\a} 
~~~~~~~~~~~~~~~~~~~~~~~\nonumber\\
A_{B12}^\a = {\cosh z_\a v \over \sinh z_\a} - {1\over z_\a}
~~~~~~~~~~~~~~~~~~~~
A_{F12}^\a = G_{F12} {\sinh z_\a v \over \cosh z_\a} ~~~~~~~~~~~~
~~~~~~~~~~~\nonumber\\
A_{B11}^\a = A_{B22}^\a = \coth z_\a - {1\over z} ~~~~~~~~~~~~~~
A_{F11}^\a = A_{F22}^\a = \tanh z_\a ~~~~~~~~~~~~~~~~~~~~\nonumber\\
\end{eqnarray*}
where $G_{F12}^2 = 1$.}
\begin{eqnarray}
C_\a ~=~ {1\over2}(1 -v^2) \Bigl(1 - {1\over12}(1 - v^2)z_\a^2 + \ldots\Bigr)
~~~~~~~~~~~~~~~~~~~~~~~~~~~\nonumber\\
{}\nonumber\\
S_{\a\b} ~=~ -(1 - v^2)\Bigl(1 - {1\over2}(1 - {1\over3}v^2)(z_\a^2 + z_\b^2)
+ \ldots \Bigr)
~~~~~~~~~~~~~~~~\nonumber\\
{}\nonumber\\
A_{\a\b} ~=~ -{1\over4} (1 - v^2)^2 z_\a z_\b + \ldots ~~~~~~~~~~~~~~~~~~~~~
~~~~~~~~~~~~~~~~~~~~
\label{eq:ds}
\end{eqnarray}

Recalling $z_+ = iesa$ and $z_- = -esb$, and noting $a+ib = 2\sqrt{\XX}$,
we verify that the combinations of the $S_{\a\b}$ and $A_{\a\b}$ occurring
in eqs.(\ref{eq:dm}) and (\ref{eq:dn}) have precisely the correct $\XX$, 
$\bar \XX$ dependence to cancel the inverse field strength factors.
Obviously this must happen to ensure that the dispersion relation involves
only positive powers of the background field strengths, but it is a very
non-trivial consistency check. For the coefficients $\AA$ and $\BB$ we find
explicitly
\begin{eqnarray}
\AA ~=~  -e^2 s^2~ \sqrt{\XX \bar\XX}~ (1-v^2)(3 - {1\over3}v^2)~+~ \ldots
\nonumber\\
\BB ~=~ e^2 s^2~ \XX~ (1-v^2)^2 ~+~\ldots ~~~~~~~~~~~~~~~~~~~~
\label{eq:dt}
\end{eqnarray}
while 
\begin{equation}
C_+ - C_- ~=~ e^2 s^2 {1\over6}(1-v^2)^2 \sqrt{\XX\bar\XX}~~~~~~~~~~~~~~~~
~~~~~~~~~~~~~~~~~~~~
\label{eq:dtadd}
\end{equation}

\vskip0.3cm
While we could continue this analysis for a general constant background field
with arbitrary $\FF$ and $\GG$, since our main interest is in the frequency
dependence of the dispersion relation we now simplify to the important 
special case of a pure magnetic field.

In this case, $\FF = {1\over2} B^2$, $\GG = 0$ so $a = \sqrt{2} \FF$,
$b=0$ and $z_+ = ies\sqrt{2\FF}$, $z_- = 0$. It is convenient to introduce
the notation $z = -i z_+ = esB$, with the hyperbolic functions of $z_+$ 
becoming trigonometric functions of $z$. Then, since $\BB$ is real, the
eigenvalues of the dispersion relation simply involve the combinations:
\footnote{These expressions can be directly compared with the classic work
on photon propagation in a magnetic field by Tsai and Erber 
\cite{Tsai:1974fa,Tsai:1975iz}.
The equivalences are:
${1\over2}(\AA + \BB) \rta - {\tan z \over z} N_{\perp}$ and 
${1\over2}(\AA - \BB) \rta - {\tan z \over z} N_{\parallel}$, eqs.(49) and
(48) of ref.\cite{Tsai:1974fa,Tsai:1975iz} respectively.  
The exponent also reproduces
the equivalent expression in ref.\cite{Tsai:1974fa,Tsai:1975iz}.}
\begin{eqnarray}
{1\over2} (\AA + \BB) ~=~ {\cos zv\over \cos z} - {v \sin zv\over\sin z}
- 2 {\cos zv - \cos z \over \cos z \sin^2 z}
~~~=~-{1\over4}(1-v^2)(1+{v^2\over3})z^2 ~+~O(z^4)
\nonumber\\
{1\over2} (\AA - \BB) ~=~ (1 - v^2) - {\cos zv\over \cos z} + 
{v \sin zv\over \sin z}~~~=~-{1\over2}(1-v^2)(1-{v^2\over3}) z^2 ~+~O(z^4)
~~~~~~~~~
\label{eq:du}
\end{eqnarray}
while the exponent involves
\begin{equation}
C_+ - C_- ~=~ -{1\over2}(1 - v^2) + {\cos zv - \cos z\over z \sin z}
~~~=~~{1\over24}(1-v^2)^2 z^2 ~+~ O(z^4)
~~~~~~~~~~~~
\label{eq:dv}
\end{equation}
\vskip0.3cm
Collecting everything, we find that the dispersion relation (\ref{eq:dp})
for a pure magnetic background field becomes:
\begin{eqnarray}
k^2 ~+~ {\a\over4\pi} \int_0^\infty {ds\over s} e^{-ism^2}
\int_{-1}^1 dv ~ {z\over \tan z}~~ 
 \exp\Bigl(-is^3 e^2 B_{\perp}^2 \w^2 ~{1\over 2z^2}(C_+ - C_-)~
\Bigr)~~~~~~~~~~~~~~~~~
\nonumber \\ 
{}\nonumber\\
\times~\Bigl[s^2 e^2 B_{\perp}^2 \w^2~ {1\over2z^2}(\AA \pm \BB)\Bigr]~~=~~0
~~~~~~~~~~
\label{eq:dw}
\end{eqnarray}

This is our final analytic result for the dispersion relation. We 
highlight three features. First, notice how the null energy combination 
$T_{\m\n}\ell^\m \ell^\n$ 
automatically projects out the relevant component of the background field, 
in this case the component orthogonal to the direction of propagation, 
i.e.~$B_{\perp}^2 = B_1^2 + B_2^2$. 

Next, note that the angle $\theta$ in eq.(\ref{eq:dq}) defining the 
direction of linear polarisation for the eigenvalues of $k^2$ is given
in this case by $\tan\theta = -{\Ima \phi_0 \over \Rea \phi_0}
= {B_1\over B_2}$. This shows that the eigenvectors $a_{\pm}^\m$ 
correspond to polarisation perpendicular (+) and parallel ($-$) to
the projection of the ${\bf B}$ field in the plane orthogonal to the
direction of propagation.

Finally, as we now show explicitly, the presence of the factor
$\w^2$ in the exponent in eq.(\ref{eq:dw}) (inherited from the
null energy term $T_{\m\n}k^\m k^\n$) means that for high frequencies
the exponent will oscillate rapidly and in fact drive the whole integral
to zero. This is the mechanism which ensures that in the high frequency
limit the QED refractive index becomes unity, ensuring $v_{\rm wf} = 1$.

\vskip0.3cm
To see this in detail, consider eq.(\ref{eq:dw}) in the weak field
approximation. The appropriate small dimensionless parameter is
$eB_{\perp}/m^2$. The dispersion relation reduces to
\begin{eqnarray}
k^2 ~-~ {\a\over8\pi} \biggl({eB_{\perp}\over m^2}\biggr)^2 \w^2~
\int_{-1}^1 dv~(1-v^2) \left\{\matrix{{1\over2}(1 + {v^2\over3})\cr {}\cr
(1 - {v^2\over3})}\right\}~~~~~~~~~~~~~~~~~~~~~~~~~~~~~~~~~~~
\nonumber\\
\times \int_0^\infty dt~t~ \exp\biggl(-i\Bigl[t + {1\over48}(1-v^2)^2  
\Bigl({eB_{\perp}\over m^2}\Bigr)^2
{\w^2\over m^2} t^3 \Bigr]\biggr) ~=~0
\label{eq:dx}
\end{eqnarray}
for the $\left\{\matrix{+\cr -\cr}\right\}$ eigenvalues respectively.
The corresponding result for the weak-field QED refractive index
is therefore
\begin{eqnarray}
n(\w) ~=~1~-~ {\a\over16\pi} \biggl({eB_{\perp}\over m^2}\biggr)^2~
\int_{-1}^1 dv~(1-v^2) \left\{\matrix{{1\over2}(1 + {v^2\over3})\cr 
{}\cr (1 - {v^2\over3})}\right\}~~~~~~~~~~~~~~~~~~~~~~~~~~~~~~
\nonumber\\
{}\nonumber\\
\times~
\int_0^\infty dt~t~ \exp\biggl(-i\Bigl[t + {1\over48}(1-v^2)^2  
\Bigl({eB_{\perp}\over m^2}\Bigr)^2
{\w^2\over m^2} t^3 \Bigr]\biggr)~~~ 
\label{eq:dy}
\end{eqnarray}
The (inverse) phase velocity and the absorption coefficient are 
determined by the real and imaginary parts of $n(\w)$.
The important observation is that the term
in the exponent proportional to $t^3$ is vital for the high frequency
behaviour of $n(\w)$. Although it is suppressed by the weak-field 
parameter $eB_{\perp}/m^2$, it cannot be dropped because it is
accompanied by the arbitrarily large $\w^2/m^2$ factor. 
This is why it is necessary to evaluate this exponential contribution
to the vacuum polarisation, rather than restrict to a simple power 
series expansion in the background field.
Indeed, this is one of the major challenges in extending this
analysis to background gravitational fields.

In the low-frequency limit, the $t$ integral is purely real and reduces to 
$\int_0^\infty dt~t~ e^{-it} = -1$. This leaves 
\begin{equation}
n(0)~=~ 1 ~+~ {\a\over90\pi} \biggl({eB_{\perp}\over m^2}\biggr)^2 
[4_+,7_-]
\label{eq:dz}
\end{equation}
in agreement with the result (\ref{eq:bjj}) derived from the
Euler-Heisenberg low-energy effective action.

In order to map out the full frequency dependence of the refractive index,
we need the following special functions defined by Airy integrals: \footnote{
We follow the notation of ref.\cite{MathWorld}.
$Gi(x)$, defined via the integral
$$
Gi(x) = {1\over\pi}\int_0^\infty dt~\sin\Bigl(xt + {t^3\over3}\Bigr)
$$
can be expressed in terms of the Airy functions $Ai(x)$ and $Bi(x)$,
or alternatively hypergeometrics, as follows:
\begin{eqnarray*}
Gi(x) ~=~ {1\over3}Bi(x) ~+~ \int_0^\infty dt~ \Bigl[
Ai(x) Bi(t) - Ai(t) Bi(x) \Bigr]~~~~~~~~~~~~~~~~~~~~~~~~~~~~~~~
~~~~~~~~~~~~~~~~ \nonumber\\
=~ {1\over3}Bi(x) -{1\over2\pi}z^2~ {}_1F_4\Bigl(1;{2\over3},{5\over6},
{7\over6},{4\over3};{1\over1296}z^6\Bigr)
-{1\over40\pi}z^5~ {}_1F_4\Bigl(1;{7\over6},{4\over3},{5\over3},{11\over6};
{1\over1296}z^6\Bigr)\nonumber\\
\end{eqnarray*}
(Note the typographical error in ref.\cite{MathWorld} where the 
factor `$z^5$' is omitted.) The prefactor in our definition (\ref{eq:daa})
of $G(x)$ compensates the asymptotic behaviour $Gi^\prime(x) \sim 
-{1\over\pi x^2} + \ldots$ as $x\rta\infty$. In eqs.(\ref{eq:dcc}) and
(\ref{eq:ddd}) the change of variable $u = (1-v^2)$ has been made 
to simplify the final formulae.} 
\begin{equation}
\int_0^\infty dt~t~\cos\Bigl[x\Bigl(t+{t^3\over3}\Bigr)\Bigr] ~~=~~ 
-{1\over x^2}~G(x^{2\over3}) ~~~~
\label{eq:daa}
\end{equation}
and
\begin{equation}
\int_0^\infty dt~t~\sin\Bigl[x\Bigl(t+{t^3\over3}\Bigr)\Bigr] ~~=~~ 
-{1\over \sqrt{3}}~ K_{2\over3}\Bigl({2\over3}x\Bigr) 
\label{eq:dbb}
\end{equation}
Here, we have defined $G(x) = -\pi x^2 {d\over dx}Gi(x)$, where the
function $Gi(x)$ is related to the standard Airy functions.

After some further simplification, and introducing the rescaled frequency
$\Omega = {eB_{\perp}\over m^2}{\w\over m}$, we find the following 
results for the real and imaginary parts of the refractive index: 
\begin{equation}
\Rea n(\Omega) ~~=~~ 1 + {\a\over\pi} \biggl({eB_{\perp}\over m^2}\biggr)^2
~{1\over24}\int_0^1 du~u~(1-u)^{-{1\over2}}~\left\{\matrix{(1-{u\over4})\cr
{}\cr (1+{u\over2})\cr}\right\}~
G\biggl(\Bigl({1\over4}u\Omega\Bigr)^{-{2\over3}}\biggr)~~~~~~~~~~~~
\label{eq:dcc}
\end{equation}
and
\begin{equation}
\Ima n(\Omega) ~~=~~ 1 + {\a\over\pi} \biggl({eB_{\perp}\over m^2}\biggr)^2
~{2\over{3\sqrt{3}}}\int_0^1 du~u^{-1}~(1-u)^{-{1\over2}}~
\left\{\matrix{(1-{u\over4})\cr {}\cr (1+{u\over2})\cr}\right\}~
\Omega^{-2}~K_{2\over3}\biggl(\Bigl({3\over8}u\Omega\Bigr)^{-1}\biggr)
\label{eq:ddd}
\end{equation}

\FIGURE
{\epsfxsize=10cm\epsfbox{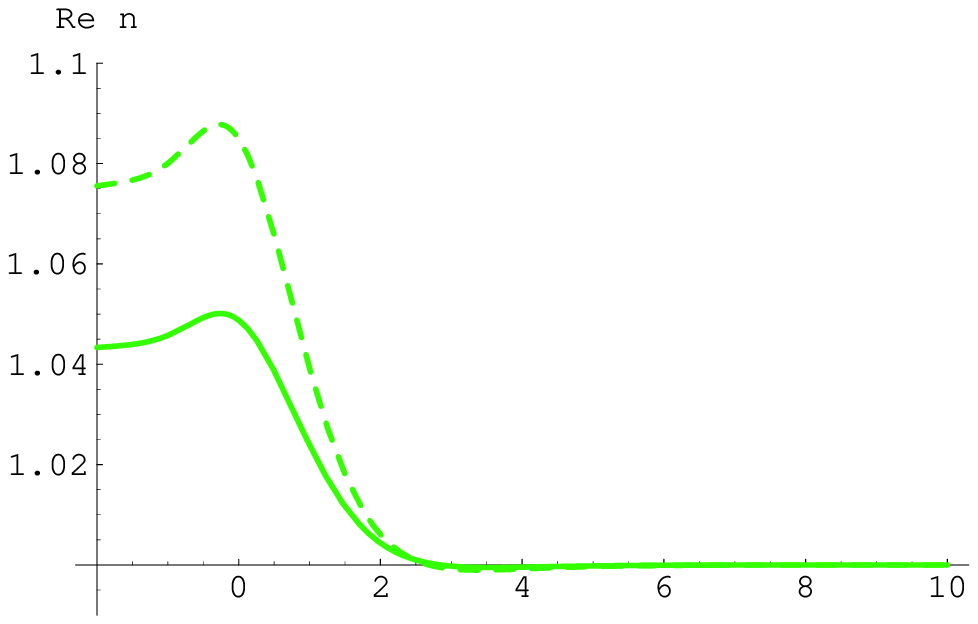}
\caption{This figure shows the real part of the QED
refractive index $\Rea n(\Omega)$ for a weak background magnetic field.
The horizontal axis $\ln \Omega$ measures frequency on a logarithmic 
scale, with $\Omega = {eB_{\perp}\over m^2}{\w\over m}$. The magnified
vertical scale corresponds to setting 
${\a\over\pi} \bigl({eB_{\perp}\over m^2}\bigr)^2 =1$ in 
eq.(\ref{eq:dcc}). The solid (dotted) line shows the 
refractive index for the polarisation $a_{+}^\m$ $(a_{-}^\m)$ 
orthogonal (parallel) to the magnetic field.}
}
These expressions are evaluated numerically and the results plotted
in Figs.~3,4,5. 
The refractive index for QED shows the same essential features as the 
single absorption band model described in section 3. The absorption 
coefficient $\Ima n(\Omega)$ due to pair creation is positive
and shows the expected single peak.
The real part $\Rea n(\Omega)$ initially rises from the values
(\ref{eq:dz}) then falls away gradually to just below 1 before approaching
the asymptotic value 1 from below, with frequency dependence 
$\Omega^{-{4\over3}}$. In terms of the phase velocity 
$v_{\rm ph}(\w) = 1/\Rea n(\w)$, the low-frequency limit is subluminal,
while asymptotically $v_{\rm ph}(\w)$ approaches $v_{\rm wf} = 1$ from
the superluminal side. 
\FIGURE
{\epsfxsize=10cm\epsfbox{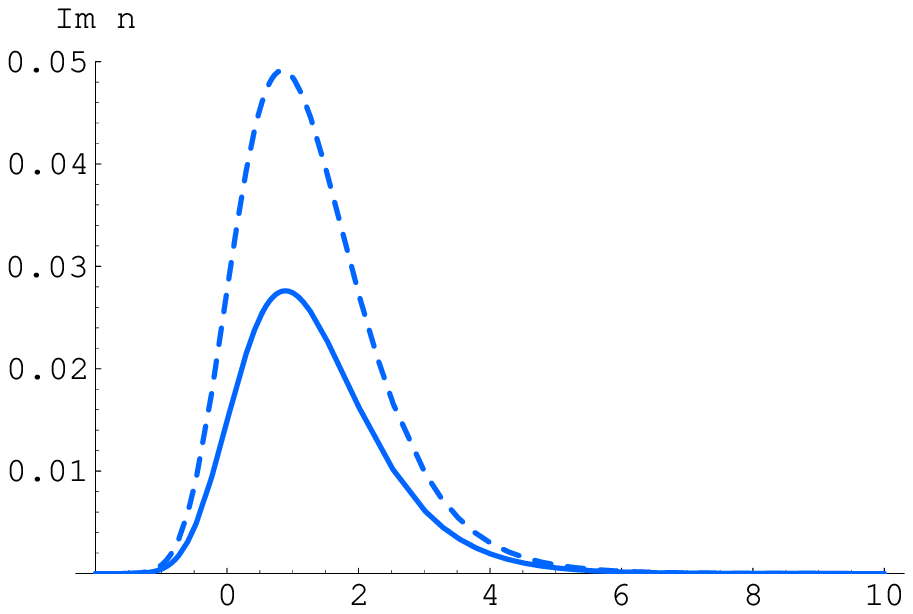}
\caption{The absorption coefficient $\Ima n(\Omega)$
plotted against $\ln\Omega$, corresponding to Fig.~3.}
}
\FIGURE
{\epsfxsize=6cm\epsfbox{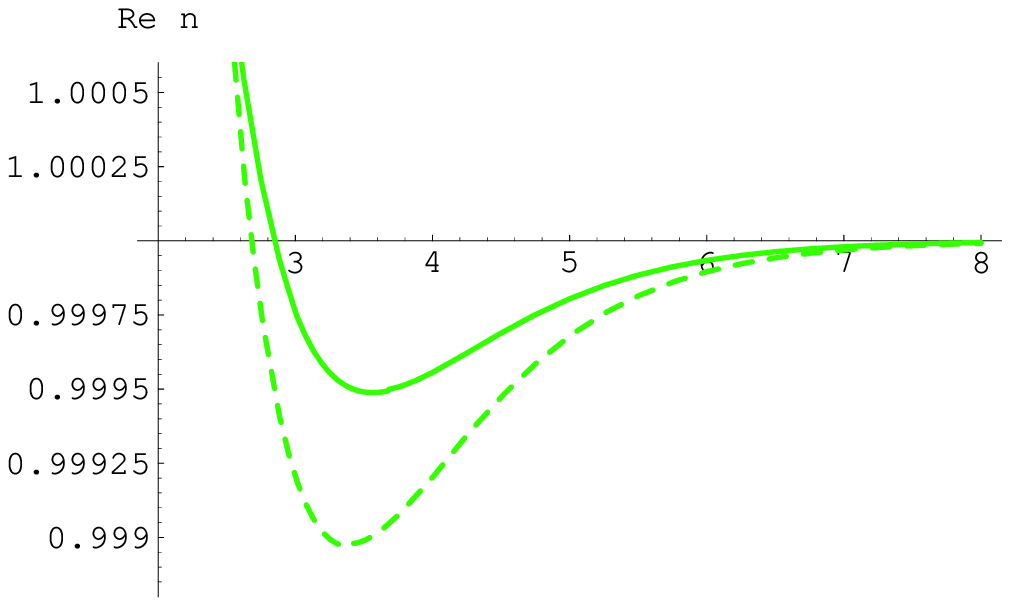}
\caption{This figure shows a magnification of the region of the refractive 
index curve where $\Rea n(\Omega)$ falls just below 1.}
}
It is worth noting that the absorption band in QED is very broad.
This means that there is an appreciable difference between the frequency
$\Omega_{\rm max} \simeq 2.5$ defining the maximum of the absorption
coefficient and the frequency $\Omega_0 \simeq 20$ where
(with the definition following the simple model (\ref{eq:ck})) the 
refractive index passes through 1. The fall-off of $\Rea n(\Omega)$ is
slow and it has only a very shallow dip into the `superluminal' regime.
(These effects are partially masked by the use of the log scale for 
$\Omega$ in the figures.)
The corresponding frequencies $\w_{\rm max}$ and $\w_0$ are inversely
proportional to the background field $B_{\perp}$.

\vskip0.3cm
We now comment briefly on the experimental implications of these results.
The most striking phenomenon is `vacuum electromagnetic birefringence'. 
A general linear polarisation is a superposition of the eigenstates
$a_{+}^\m$ and $a_{-}^\m$ which propagate with different phase velocities.
After propagating through some distance, this velocity difference
introduces a phase difference between the eigenstates, and the resulting
wave then describes elliptic polarisation with the principal axis of the
ellipse rotated relative to the initial direction of linear 
polarisation.\footnote{The PVLAS collaboration 
\cite{Zavattini:2005tm, Zavattini:2005ca}
has recently reported the observation of a rotation of the polarisation
plane for linearly polarised light in a 5.5T magnetic field using
their high-sensitivity optical ellipsometer. The origin of this
observation is not yet understood. The vacuum electromagnetic birefringence
described here is negligible in this system, prompting the speculation
that the effect may be due to the existence of a new ultra-light 
axion-like particle with mass in the region of $10^{-3}$eV.}

Experiments to look for vacuum electromagnetic birefringence are currently
being planned \cite{Heinzl:2006xc,Heinzl:2006pn} using either an X-ray 
free-electron laser or a Thomson backscattering source to provide a 
high-frequency polarised pulse to scatter from a high-intensity laser 
background. Even for such high frequencies, however, we still only have 
${\w\over m}\sim 10^{-2}$ and the attainable background fields
from petawatt lasers with intensity around $10^{22}~{\rm Wcm}^{-2}$ 
correspond to $eB_{\perp}/m^2 \sim 10^{-4}$. This means that 
currently envisageable experiments will only be sensitive to the
low-frequency sector of the weak-field QED refractive index in Fig.~3,
i.e.~the region of normal dispersion (rising $\Rea n(\w)$).
The anomalous dispersion (falling $\Rea n(\w)$) region correlated with
the peak of the absorption coefficient $\Ima n(\w)$
remains beyond the direct reach of current technology.

\vskip0.5cm
To conclude this section, we have shown in explicit detail how the
KK dispersion relation (\ref{eq:cj}) controls the relation between
the IR and UV limits of the phase velocity in a well-understood quantum
field theory. The absorption coefficient $\Ima n(\w)$ due to pair
creation is positive, in accordance with expectations from the
optical theorem, ensuring that the low-frequency phase velocity 
$v_{\rm ph}(0)$ is less than its UV limit $v_{\rm ph}(\infty) = v_{\rm wf}$.
Causality requires $v_{\rm wf}\le 1$ and as expected we find that
$v_{\rm wf} = 1$ in QED, a result which arises technically through the
appearance of a rapidly-oscillating exponential factor in the vacuum
polarisation. It therefore follows that in QED there is indeed a 
causality bound $v_{\rm ph}(0) < 1$. As we have seen, the IR limit 
of the phase velocity is determined by a combination of the null
projection of the energy momentum tensor and the coefficients of the 
leading irrelevant operators in the low-energy (Euler-Heisenberg) 
effective action. It finally follows that, under the assumption that 
the null energy condition $T_{\m\n}k^\m k^\n \ge 0$ holds, causality 
does impose the positivity bound (\ref{eq:bhh}) on these parameters.

\section{Superluminality and QED in gravitational fields}

We now consider photon propagation in a background gravitational field,
taking into account the quantum corrections induced by vacuum polarisation.
Remarkably, in this case the low-frequency phase velocity $v_{\rm ph}(0)$
can become superluminal.

The low-energy effective action for QED in a background gravitational
field (the analogue of the Euler-Heisenberg action) was first calculated
by Drummond and Hathrell in the original paper \cite{Drummond:1979pp}
in which they discovered the possibility of vacuum-polarisation 
induced superluminal propagation. Integrating out the electron field, 
the one-loop effective action is\footnote{There is a further term, 
$d D_\m F^{\m\l} D_\n F^\n{}_\l$ with $d = -24$, 
but this only affects the dispersion relation at $O(\a k^2)\sim O(\a^2)$ 
since its contribution to the Maxwell equation involves the equation of motion 
factor $D_\m F^{\m\l}$ which is itself already of $O(\a)$. 
The Ricci scalar term also does not contribute to the dispersion
relation at $O(\a)$ for the same reason.}
\begin{eqnarray}
\int dx\sqrt{-g}~ \LL ~~=~~ \int dx\sqrt{-g}\biggl[-{1\over4}F_{\m\n}F^{\m\n}
~~~~~~~~~~~~~~~~~~~~~~~~~~~~~~~~~~~~~~~~~~~~~~~~~~~~\nonumber\\
+ {\a\over720\pi}{1\over m^2}\Bigl(a R F_{\m\n}F^{\m\n} 
+ b R_{\m\n}F^{\m\l}F^\n{}_\l
+ c R_{\m\n\l\r}F^{\m\n}F^{\l\r}\Bigr)
\biggr]
\label{eq:ea}
\end{eqnarray}
The coefficients of the leading irrelevant operators of type $\RR F F$
are $a = -5$, $b = 26$ and $c = -2$.

To study photon propagation with this effective action, we again use the
geometric optics methods described in section 2, making the additional
assumption that the background field is slowly-varying on 
scales comparable to the wavelength of light. In contrast to the 
electromagnetic backgrounds, we cannot restrict to constant gravitational 
fields since for a spacetime of constant curvature\footnote{A spacetime of 
constant curvature is characterised by
$$
R_{\m\n} = {1\over4} R g_{\m\n}~~~~~~~~~~~~
R_{\m\n\l\r} = {1\over12}R (g_{\m\l}g_{\n\r} - g_{\m\r}g_{\n\l})
$$
} the only independent component is the Ricci scalar, which does not affect 
the dispersion relation at $O(\a)$. In this case, omitting the terms
involving derivatives of the curvatures, which are suppressed by
$O({\l\over L})$ for wavelength $\l$ and curvature scale $L$, the modified 
Maxwell equation gives rise to the following relation (the analogue of 
eq.(\ref{eq:bq}) for a background electromagnetic field):
\begin{equation}
k^2 a^\n - k.a k^\n- {\a\over 720\pi}{1\over m^2} \Bigl(
2b R_{\l\r}(k^\l k^\r a^\n - k^\l k^\n a^\r) -
8c R^\n{}_{\l\m\r} k^\l k^\r a^\m \Bigr) ~=~ 0
\label{eq:eb}
\end{equation}

We now re-express this in terms of the Ricci and Weyl tensors and introduce
the NP formalism. In an analogous way to the electromagnetic field strengths,
the ten independent components of the Weyl tensor are represented by five
complex scalars $\Psi_i$ ($i = 0,\ldots 4$), where \cite{Chandra}
\begin{eqnarray}
&\Psi_0 &= - C_{\m\n\l\r}\ell^\m m^\n \ell^\l m^\r \nonumber\\
&\Psi_1 &= - C_{\m\n\l\r}\ell^\m n^\n \ell^\l m^\r \nonumber\\
&\Psi_2 &= - C_{\m\n\l\r}\ell^\m m^\n \bar m^\l n^\r \nonumber\\
&\Psi_3 &= - C_{\m\n\l\r}\ell^\m n^\n \bar m^\l n^\r \nonumber\\
&\Psi_4 &= - C_{\m\n\l\r}n^\m \bar m^\n n^\l \bar m^\r
\label{eq:ec}
\end{eqnarray}
The independent components of the Ricci tensor are described by
four real and three complex scalars:
\begin{eqnarray}
&\Phi_{00} &= -{1\over2}R_{\m\n} \ell^\m \ell^\n 
~~~~~~~~~~~~~~~~~~~~~~~~~~~~~~~~~~~~~~~~~
\Phi_{01} = \Phi_{10}^* = -{1\over2}R_{\m\n} \ell^\m m^\n
~~~~~~~~~~~~~~~~\nonumber\\
&\Phi_{11} &= -{1\over4}R_{\m\n} (\ell^\m n^\n + m^\m \bar m^\n) 
~~~~~~~~~~~~~~~~~~~~~~~~~~
\Phi_{02} = \Phi_{20}^* = -{1\over2}R_{\m\n} m^\m m^\n
~~~~~~~~~~~~~~\nonumber\\
&\Phi_{22} &= -{1\over2}R_{\m\n} n^\m n^\n 
~~~~~~~~~~~~~~~~~~~~~~~~~~~~~~~~~~~~~~~~
\Phi_{12} = \Phi_{21}^* = -{1\over2}R_{\m\n} n^\m m^\n
~~~~~~~~~~~~~~~\nonumber\\
&\Lambda  &= {1\over24}R = 
{1\over12} R_{\m\n}(\ell^\m n^\n - m^\m \bar m^\n) 
\label{eq:ed}
\end{eqnarray}

As before, we take $k_\m = \sqrt{2}\w \ell^\m$ and the polarisation vectors 
$a^\m = \a m^\m + \b \bar m^\m$. Then, eq.(\ref{eq:eb}) becomes
\begin{equation}
\left(\matrix{k^2 +{\a\over90\pi}{\w^2\over m^2}(b+2c)\Phi_{00}
& -{\a\over90\pi}{\w^2\over m^2} 2c \Psi_0^* \cr
{}\cr
-{\a\over90\pi}{\w^2\over m^2} 2c \Psi_0 
& k^2 + {\a\over90\pi}{\w^2\over m^2}(b+2c)\Phi_{00} \cr}\right)
\left(\matrix{\a\cr {}\cr \b\cr}\right)~~=~~ 0 
\label{eq:ee}
\end{equation}
where we have used the important identity 
$C_{\m\n\l\r}\ell^\m m^\n \ell^\l \bar m^\r = 0.$
The eigenvalues give the new light cone
\begin{equation}
k^2 ~+~ {\a\over90\pi}{\w^2\over m^2}\Bigl[(b+2c) \Phi_{00} ~\pm~
2c \sqrt{\Psi_0 \Psi_0^*}~\Bigr]~=~0
\label{eq:ef}
\end{equation}
The polarisation eigenstates are
\begin{eqnarray}
&a^\m_+ ~&=~{1\over\sqrt{2}} (e^{i\theta} m^\m + e^{-i\theta}\bar m^\m)
\nonumber \\
&a^\m_- ~&=~-{i\over\sqrt{2}} (e^{i\theta} m^\m - e^{-i\theta}\bar m^\m)
\label{eq:eg}
\end{eqnarray}
with the phase defined by $\Psi_0^* = |\Psi_0|e^{2i\theta}$. 

The modified light cone is therefore
\begin{equation}
k^2 ~+~ {\a\over360\pi}{1\over m^2}\Bigl[-(b+2c)~ R_{\m\n}k^\m k^\n 
~\pm~ 4c ~ |C_{\m\n\l\r}k^\m m^\n k^\l m^\r| \Bigr] ~=~0
\label{eq:eh}
\end{equation}
corresponding to a phase velocity
\begin{equation}
v_{\rm ph}(0) ~=~ 1  -{\a\over360\pi} {1\over m^2} 
\Bigl[-(b+2c) ~R_{\m\n}\ell^\m \ell^\n ~\pm~ 
4c~|C_{\m\n\l\r} \ell^\m m^\m \ell^\n  m^\r| \Bigr]
\label{eq:ei}
\end{equation}

Notice immediately that using the Einstein equation, the Ricci tensor
can be re-expressed in terms of the energy-momentum tensor by $R_{\m\n}
= 8\pi T_{\m\n}$ (in $G=1$ units) and the first term can be written
in familiar form involving the null energy projection $T_{\m\n}k^\m k^\n$.
In the gravitational case, however, the phase velocity also has
a new, polarisation-dependent contribution involving the Weyl
curvature. The relation between superluminality and the null energy
condition is therefore more subtle for QED in a gravitational background
field. 

For Weyl-flat spacetimes, the situation is similar to the electromagnetic
case. Assuming the null-energy condition holds, the sign of 
$R_{\m\n}k^\m k^\n$ is fixed and the question of whether $v_{\rm ph}(0)$
is superluminal is determined by the coefficients of the leading
irrelevant operators in the low-energy effective action. Specifically,
a subluminal $v_{\rm ph}(0) < 1$ requires a bound $(b+2c) < 0$.
Remarkably, this is {\it violated} by QED in a background curved spacetime
where, as we have seen, $b = 26$ and $c = -2$.\footnote{Note that 
${1\over2}(b+2c) = 11$, revealing the universal factor 11 first
identified by Latorre {\it et al.} in ref.\cite{Latorre:1994cv} 
as characterising
`energy density' effects on the speed of light. Note that the polarisation
sum $\sum_{+,-} v_{\rm ph}(0)$ for electromagnetic fields involves the 
coefficients $4 + 7 = 11$. This universality arises because of the common 
dependence on the one-electron-loop vacuum polarisation.} Both polarisations 
have the same phase velocity and there is no birefringence.

For Ricci-flat spacetimes, i.e.~vacuum solutions of Einstein's equation
where $T_{\m\n} = 0$, eq.(\ref{eq:ei}) shows that for some directions
and polarisations, photons must propagate with a superluminal 
$v_{\rm ph}(0) > 1$ {\it independent} of the values of the low-energy
couplings. If one polarisation is subluminal, the other is necessarily
superluminal, satisfying the polarisation sum rule $\sum_{+,-} \d 
v_{\rm ph}(0) = 0$. The theory exhibits {\it vacuum gravitational 
birefringence.}

We illustrate this with two examples. First, consider a 
Friedmann-Robertson-Walker (FRW) spacetime 
\cite{Drummond:1979pp,Shore:2003jx}. 
This has vanishing Weyl tensor, is spatially isotropic, and has Ricci 
curvature corresponding to the energy-momentum tensor 
$T_{\m\n} = (\r + P) e^t_\m  e^t_\n - P g_{\m\n}$, where $\r$
is the energy density, $P$ is the pressure and $e^t_\m$ 
specifies the time direction in a comoving orthonormal frame.
The speed of light is the same in all directions and is polarisation
independent. Eqs.(\ref{eq:eh}), (\ref{eq:ei}) give
\begin{equation}
k^2 ~=~ {22\over45}{\a\over m^2}~ T_{\m\n}k^\m k^\n 
\label{eq:ej}
\end{equation}
and
\begin{equation}
v_{\rm ph}(0) ~=~ 1 ~+~ {11\over45}{\a\over m^2}~(\r + P)
\label{eq:ek}
\end{equation}
This confirms the surprising result that the {\it gravitational} effect
of a positive null energy projection $T_{\m\n}k^\m k^\n$ is to increase
the phase velocity, resulting in superluminal propagation in Weyl-flat
spacetimes.

As an example of a Ricci-flat background with non-vanishing Weyl tensor,
consider Schwarzschild spacetime. In the NP formalism, a standard choice
of null tetrad $\ell^\m, n^\m, m^\m, \bar m^\m$ is made which reflects 
the properties of the null geodesics. Specifically, in terms of a 
conventional orthonormal tetrad $(e^\m_t, e^\m_r, e^\m_\theta, e^\m_\phi)$,
we assign
\begin{eqnarray}
&\ell^\m &= {1\over\D}(r^2,\D,0,0) \nonumber\\
&n^\m &= {1\over2r^2}(r^2,-\D,0,0) \nonumber\\
&m^\m &= {1\over \sqrt{2} r}(0,0,1,i~{\rm cosec}\theta)
\label{eq:el}
\end{eqnarray}
where $\D = r^2 - 2Mr$. We restrict ourselves to planar motion, so without
loss of generality set $\theta = \pi/2$ from now on.

To consider modifications to the principal null geodesics (radial directed)
we choose $k^\m = \sqrt{2}\w \ell^\m$. The derivation of the new light 
cones follows precisely eqs.(\ref{eq:ee}) to (\ref{eq:eg}) and we find
\begin{equation}
k^2 ~\pm~ {\a\over45\pi}{\w^2\over m^2} c~|\Psi_0| ~=~ 0
\label{eq:em}
\end{equation}
Now, it is a crucial feature of Schwarzschild spacetime (and indeed a large 
class of familiar black hole spacetimes such as Kerr) that it is of Petrov 
type D and the only non-vanishing NP Weyl scalar is $\Psi_2 = -M/r^3$.
This is a consequence of the Goldberg-Sachs theorem and reflects the
shear-free character of the principal null geodesics.  It follows 
immediately that for radial motion, the light cone $k^2 = 0$ remains
unchanged and $v_{\rm ph}(0) = 1$.

Next, consider the modified null geodesics in the $r,\phi$ plane.
Expressing the tangent vector $L^\m = {dx^\m\over ds}$ in terms of the
NP basis gives\footnote{The Schwarzschild metric is 
$$
ds^2 = {\D\over r^2}dt^2 - {r^2\over\D}dr^2 - r^2(d\theta^2 + 
\sin^2\theta d\phi^2)
$$ 
The $t,\phi$ geodesic equations in the 
$\theta = {\pi\over2}$ plane are
$$
{d^2t\over ds^2} + \Bigl({\D'\over\D} -{2\over r}\Bigr) {dt\over ds}
{dr\over ds}~=~0~~~~~~~~~~~~~~~~~~
{d^2\phi\over ds^2} + {2\over r} {d\phi\over ds}{dr\over ds}~=~0
$$
and integrate to give
$$
{dt\over ds} = {r^2\over\D} ~~~~~~~~~~~~~~~~~~~~~~~~~~~~~~~~~~~~~
{d\phi\over ds} = {1\over r^2}D
$$
Substituting into the metric, we then find that for a null interval 
$$
{dr\over ds} = \Bigl(1 - {D^2\D\over r^4}\Bigr)^{1\over2}
$$
The stated result for $L^\m = {dx^\m\over ds}$ follows immediately.
The radial geodesic is the special case where the impact parameter
vanishes, $D=0$.}
\begin{equation}
L^\m ~=~ \Bigl({r^2\over\D}, F, 0, 
{D\over r^2} \Bigr)  ~~=~~ {1\over2}(1+F)\ell^\m + {r^2\over\D}(1-F)n^\m
- {i\over\sqrt{2}} {D\over r} (m^\m - \bar m^\m)
\label{eq:en}
\end{equation} 
where $D$ is the impact parameter and we have abbreviated 
$F = \Bigl(1 - {D^2 \D\over r^4}\Bigr)^{1\over2}$.
A convenient choice for the orthogonal circular polarisation vectors is
\begin{equation}
M^\m ~=~ {1\over \sqrt{2} r} (0, {D\D\over r^2}, i, -F) ~~=~~
{1\over\sqrt{2}}{D\over r}\Bigl({\D\over 2r^2}\ell^\m - n^\m \Bigr)
+ {i\over2}(1+F) m^\m + {i\over2}(1-F)\bar m^\m
\label{eq:eo}
\end{equation}
where $L.M = 0, ~ M.\bar M = -1$. Now set $k^\m = \sqrt{2}\w L^\m$
and take the polarisation eigenstates as $a^\m = \a M^\m + \b \bar M^\m$.
As in eq.(\ref{eq:ee}) we find
\begin{equation}
\left(\matrix{
k^2 -{\a\over90\pi}{\w^2\over m^2}2c~C_{LML\bar M}
&-{\a\over90\pi}{\w^2\over m^2}2c~C_{L\bar M L \bar M} 
\cr {}\cr
-{\a\over90\pi}{\w^2\over m^2}2c~C_{L M L M}
&k^2 -{\a\over90\pi}{\w^2\over m^2}2c~C_{L \bar M L M}
\cr}\right)
\left(\matrix{\a\cr {}\cr \b \cr}\right) ~~=~~0
\label{eq:ep}
\end{equation}
where $C_{LML\bar M} = C_{\m\n\l\r}L^\m M^\n L^\l \bar M^\r$ etc.
The Weyl tensor for Schwarzschild may be written as
\begin{equation}
C_{\m\n\l\r} ~=~ \Psi_2\Bigl( -[\ell_\m n_\n \ell_\l n_\r]
-[m_\m \bar m_\n m_\l \bar m_\r] + [\ell_\m m_\n n_\l \bar m_\r]
+ [\ell_\m \bar m_\n n_\l m_\r]\Bigr)
\label{eq:eq}
\end{equation}
where the notation $[\ldots]$ denotes a sum over permutations with the
symmetries of $C_{\m\n\l\r}$.
Evaluating the required components, we readily find 
\begin{equation}
C_{LML\bar M} ~=~ 0 ~~~~~~~~
C_{LMLM} ~=~ C_{L\bar M L \bar M} ~=~ -{3\over r^2}~D^2~\Psi_2 
\label{eq:er}
\end{equation}
It follows that the new light cone is
\begin{equation}
k^2 ~\pm~ {\a\over90\pi}{\w^2\over m^2}2c~ {3M\over r^5}~D^2 ~~=~~ 0
\label{eq:es}
\end{equation}
that is
\begin{equation}
v_{\rm ph}(0) ~~=~~ 1  ~
\pm~ {2\a\over45\pi}{1\over m^2}~ {3M\over r^5}~D^2
\label{eq:et}
\end{equation}
restoring $c=-2$. Since $\Psi_2$ is real, the corresponding polarisation 
eigenstates are simply
\begin{eqnarray}
&a^\m_+ ~&=~ {1\over\sqrt{2}} (M^\m + \bar M^\m) \nonumber\\
&a^\m_- ~&=~ -{i\over\sqrt{2}} (M^\m - \bar M^\m) 
\label{eq:eu}
\end{eqnarray}
$a^\m_+$ is the linear polarisation transverse to the direction of $k^\m$ 
in the plane of motion, while $a^\m_-$ is the orthogonal linear
polarisation along $e_\theta$. This provides an important example of 
gravitational birefringence with a superluminal $v_{\rm ph}(0)$. 

\vskip0.3cm
This brings us to the key question. If it is possible to have a 
superluminal phase velocity in the low-frequency limit, how can this
be reconciled with causality, given the KK dispersion relation?

First, we review briefly what is known directly about dispersion
for photon propagation in background gravitational fields.
Adapting the pioneering background field calculations of Barvinsky
{\it et al.} \cite{BGVZone,BGVZtwo},
we constructed the QED effective 
action at $O(\RR F F)$ to all orders in derivatives 
\cite{Shore:2002gw,Shore:2002gn}. This action is\footnote{
This expression is slightly simplified from the form quoted in
refs.\cite{Shore:2002gw,Shore:2002gn}. Here, we have used the identity
$$
\int dx \sqrt{-g}~R_{\m\n}D^\m D^\n F^{\l\r} F_{\l\r} ~=~
-\int dx \sqrt{-g}~R_{\m\n} D^\m F^{\l\r} D^\n F_{\l\r} ~+~
{1\over4}\int dx \sqrt{-g}~D^2 R F^{\l\r}F_{\l\r}
$$
to relate two terms considered independent in 
refs.\cite{Shore:2002gw,Shore:2002gn}. The form factors are adapted
accordingly. We have also omitted a term $D_\m F^{\m\l}\overrightarrow d_0
D_\n F^\n{}_\l$ since, as explained above, it does not contribute
to the dispersion relation at $O(\a)$. 
}
\begin{eqnarray}
\int dx\sqrt{-g}~\LL~~ =~~ \int dx \sqrt{-g}~ \biggl[ 
-{1\over4}F_{\m\n}F^{\m\n}~~~~~~~~~~~~~~~~~~~~~~~~~~~~~~~~~~~~~~~~~~~~~~
~~~~~~~~\nonumber\\
+~{1\over m^2}
\Bigl(\orta{a_0}~ R F_{\m\n} F^{\m\n}~ 
+~\orta{b_0}~ R_{\m\n} F^{\m\l}F^\n{}_{\l}~
+~\orta{c_0}~ R_{\m\n\l\r}F^{\m\n}F^{\l\r} \Bigr)~~~~ \nonumber\\
~~+~{1\over m^4}\Bigl(\orta{a_1}~ R D_\m F^{\m\l} D_\n F^\n{}_{\l}~  
+~\orta{b_1}~ R_{\m\n} D_\l F^{\l\m}D_\r F^{\r\n} ~~~~~~~~~~~~~~~~~
\nonumber\\
+~\orta{b_2}~ R_{\m\n} D^\m F^{\l\r}D^\n F_{\l\r}~
+~\orta{b_3}~ R_{\m\n} D^\m D^\l F_{\l\r} F^{\r\n}~~~~~~
\nonumber\\
~+~\orta{c_1}~ R_{\m\n\l\r} D_\s F^{\s\r}D^\l F^{\m\n} ~\Bigr)
~~~~~~~~~~~~~~~~~~~~~~\nonumber\\
+~{1\over m^6}~\orta{b_4}~ R_{\m\n} D^\m D_\l F^{\l\s}D^\n 
D_\r F^\r{}_\s   ~~\biggr]~
\nonumber\\ 
\label{eq:ev}
\end{eqnarray}
In this formula, the $\orta{a_n}$, $\orta{b_n}$, 
$\orta{c_n}$ are known form factor functions of three operators, i.e.
\begin{equation}
\orta{a_n} \equiv a_n\Bigl({D_{(1)}^2\over m^2}, {D_{(2)}^2\over m^2}, 
{D_{(3)}^2\over m^2}\Bigr)
\label{eq:ew}
\end{equation}
where the first entry ($D_{(1)}^2$) acts on the first following term
(the curvature), etc. It reduces to the Drummond-Hathrell action 
(\ref{eq:ea}) in the low-energy limit.

This action contains all the information required to extend the 
dispersion relation from the zero-frequency limit into the low-frequency 
region. Extracting this from the action involves a number of subtleties
described in full in ref.\cite{Shore:2002gn}. The result is an expression 
resembling eq.(\ref{eq:eh}), viz.
\begin{equation}
k^2 ~+~ {\a\over360\pi}{1\over m^2}\Bigl[F\Bigl({k.D\over m^2}\Bigr)
~ R_{\m\n}k^\m k^\n 
~\pm~ G\Bigl({k.D\over m^2}\Bigr) ~ |C_{\m\n\l\r}k^\m m^\n k^\l m^\r| 
\Bigr] ~=~0
\label{eq:ex}
\end{equation}
where the constant coefficients are replaced by functions of the
operator $k.D$, which describes the variations of the curvature along
the unperturbed null geodesics. The precise form of the functions $F$ 
and $G$ in terms of the form factors in eq.(\ref{eq:ev}) is given in 
ref.\cite{Shore:2002gn}.
 
However, it seems unlikely that this will be sufficient to describe
the high-frequency behaviour of the dispersion relation. An equivalent
approach applied to electrodynamics, in which we keep all orders in
derivatives but restrict to lowest order in the field strengths, would
miss the crucial exponent factor in the vacuum polarisation. This arises
naturally in both the heat kernel and worldline path 
integral approaches and it would be extraordinary if the gravitational
case was very different. This leads us to conjecture the following
form for the dispersion relation for QED in a gravitational background,
analogous to eqs.(\ref{eq:dp}) and (\ref{eq:dw}) for the electromagnetic 
case:
\begin{equation}
k^2 ~+~ {\a\over\pi} \int_0^\infty ds~ {\cal N}(s,\RR)~ 
\exp\Bigl[-is\Bigl(1 + 
s^2 \Omega^2 {\cal P}(s,\RR)\Bigr)\Bigr] ~~=~~0
\label{eq:ey}
\end{equation}
Here, $\Omega \sim {\RR\over m^2}{\w\over m}$, where $\RR$ denotes 
some generic curvature component, and ${\cal N}(s,\RR)$ and 
${\cal P}(s,\RR)$ can be expanded in powers of curvatures and derivatives,
with appropriate powers of $s$. Neglecting the exponent factor is only
valid for small $\Omega$, i.e.~for wavelengths satisfying $\l \gg
{\l_c^3\over L^2}$, where $\l_c$ is the electron Compton wavelength
and $L$ is a typical curvature scale. This is too near the IR to see
the interesting structure in the refractive index.

Assuming this picture is correct, we will not be able to complete a
direct evaluation of the full frequency dependence of the refractive
index for QED in gravitational fields until new QFT techniques are
developed which allow a calculation of the exponent contribution to 
the vacuum polarisation. The gravitational case has two main extra 
difficulties over pure electromagnetism -- one is simply the plethora
of indices associated with higher powers of curvature, but more
importantly we cannot restrict to constant background fields since,
as explained above, constant curvature spacetimes do not have the
required spacetime anisotropy needed to modify the dispersion relation.
The most promising approach would seem to involve formulating the
heat kernel or worldline path integral methods using the NP formalism
from the outset. For example, this would allow us to study black hole
backgrounds in terms of only one non-vanishing NP curvature scalar
$\Psi_2$. Even so, the analysis looks far from straightforward.

\vskip0.3cm
Now return to the KK dispersion relation (\ref{eq:cj}). Recall that this
predicts, under the usual assumption $\Ima n(\w) > 0$, that the
high-frequency limit $v_{\rm wf} = v_{\rm ph}(\infty)$ of the  
phase velocity is greater than $v_{\rm ph}(0)$. However, for QED
in gravitational fields, we have seen that it is possible to have
a superluminal low-frequency limit $v_{\rm ph}(0) > 1$. 
The question is, how can this be reconciled with causality?

We now consider the possible resolutions in turn:

\noindent 1.~~It is possible that the KK dispersion relation does not hold 
in curved spacetime. However, the assumptions in the derivation of 
eq.(\ref{eq:cj}) and similar relations for scattering amplitudes are 
the just the fundamental axioms of QFT, especially local Lorentz invariance 
and micro-causality, i.e.~the vanishing of commutators of local operators 
for spacelike separations. These should also apply in curved spacetime, 
certainly for globally hyperbolic spacetimes (which admit a foliation 
into spacelike hypersurfaces), and we have examples of superluminal 
$v_{\rm ph}(0)$ in such spacetimes. Invalidity of the KK dispersion 
relation itself therefore looks unlikely.

\noindent 2.~~It may be that $v_{\rm ph}(\infty)$ is indeed perturbatively 
greater than 1 and the physical light cone lies outside the geometric one, 
but that nevertheless the spacetime satisfies the criteria for stable 
causality with respect to the {\it effective metric} characterising the 
physical light cone. This is the situation discussed in section 3. However, 
this seems unlikely. If our conjecture (\ref{eq:ey}) about the form of the 
vacuum polarisation is correct, it seems extremely probable that 
in the high-frequency limit the presence of the factor $\Omega^2$ 
will result in the exponent term oscillating rapidly and, just as
in the electromagnetic case, driving the entire integral to zero.
The QFT evidence appears to point to $v_{\rm ph}(\infty) = 1$.

\noindent 3.~~The potential paradox would be resolved if it is possible
to have $\Ima n(\w) < 0$ in gravity. At first sight, this appears to
be ruled out by the usual QFT identification of the imaginary part of
a forward scattering amplitude with the total cross-section via the
optical theorem. Indeed, the imaginary part of the vacuum polarisation
should be related, just as in the case of a background electromagnetic 
field, to $e^+ e^-$ pair creation, which is certainly absorptive. 
Another important physical process, photon splitting in a background field
\cite{Adler:1970gg,Adler:1971wn} is also associated with $\Ima n(\w)$
positive.

It is hard to see a way out. One line of thinking, as we explore further in
the next section, is that a negative $\Ima n(\w)$ corresponds to gain,
i.e.~an amplification of the amplitude and intensity of the light wave,
rather than absorption. This may give us some hope that gravity is 
special, due to its ability to focus (see also ref.\cite{Dolgov}). 
Recalling the geometric optics
equation (\ref{eq:bf}) for the variation of the amplitude $A$ along
a geodesic, we readily see that including the $O(\a)$ contribution
from the vacuum polarisation in the low-energy effective action 
(\ref{eq:ea}) gives schematically
\begin{equation}
k^\m D_\m(\ln A) = -{1\over\sqrt{2}}~\w~\theta ~+~ 
\a ~k^\m D_\m \RR
\label{eq:ez}
\end{equation}
The optical scalar $\theta = \ell^\m{}_{;\m}$ is the expansion coefficient 
(we have set $k^\m = \sqrt{2}\w\ell^\m$ here), so for focusing geodesics
($\theta < 0$) the amplitude is increased. This is a purely classical effect.
However, there is an additional $O(\a)$ quantum contribution depending on
the change in the curvature along a geodesic. This arises from the vacuum
polarisation and will in general show a frequency dependence. The expansion
scalar $\theta$ and associated Raychaudhuri equation equally show
$O(\a)$ corrections \cite{Ahmadi:2006nq}. It seems natural that the
$O(\a)$ corrections to the geodesics induced by the vacuum polarisation
should be able to enhance focusing as well as
divergence, corresponding to amplification as well as attenuation.
This would suggest that a negative $\Ima n(\w)$ may after all be a viable
possibility in gravity.

Overall, however, the situation remains puzzling. Obviously what is needed
to replace speculation by hard physics is a direct calculation of the
momentum dependence of the vacuum polarisation for QED in a curved
spacetime. This challenging technical problem remains open.

\section{Laser-atom interactions in $\Lambda$-systems:~EIT and 
Raman gain lines}

In this section, we develop some intuition on how it may be possible to
have $v_{\rm ph}(\infty) < v_{\rm ph}(0)$ and $\Ima n(\w) < 0$ by
studying light propagation in an atomic physics context. Specifically,
we consider so-called `$\L$-systems', illustrated in Fig.~6.

\FIGURE
{\epsfxsize=6.8cm\epsfbox{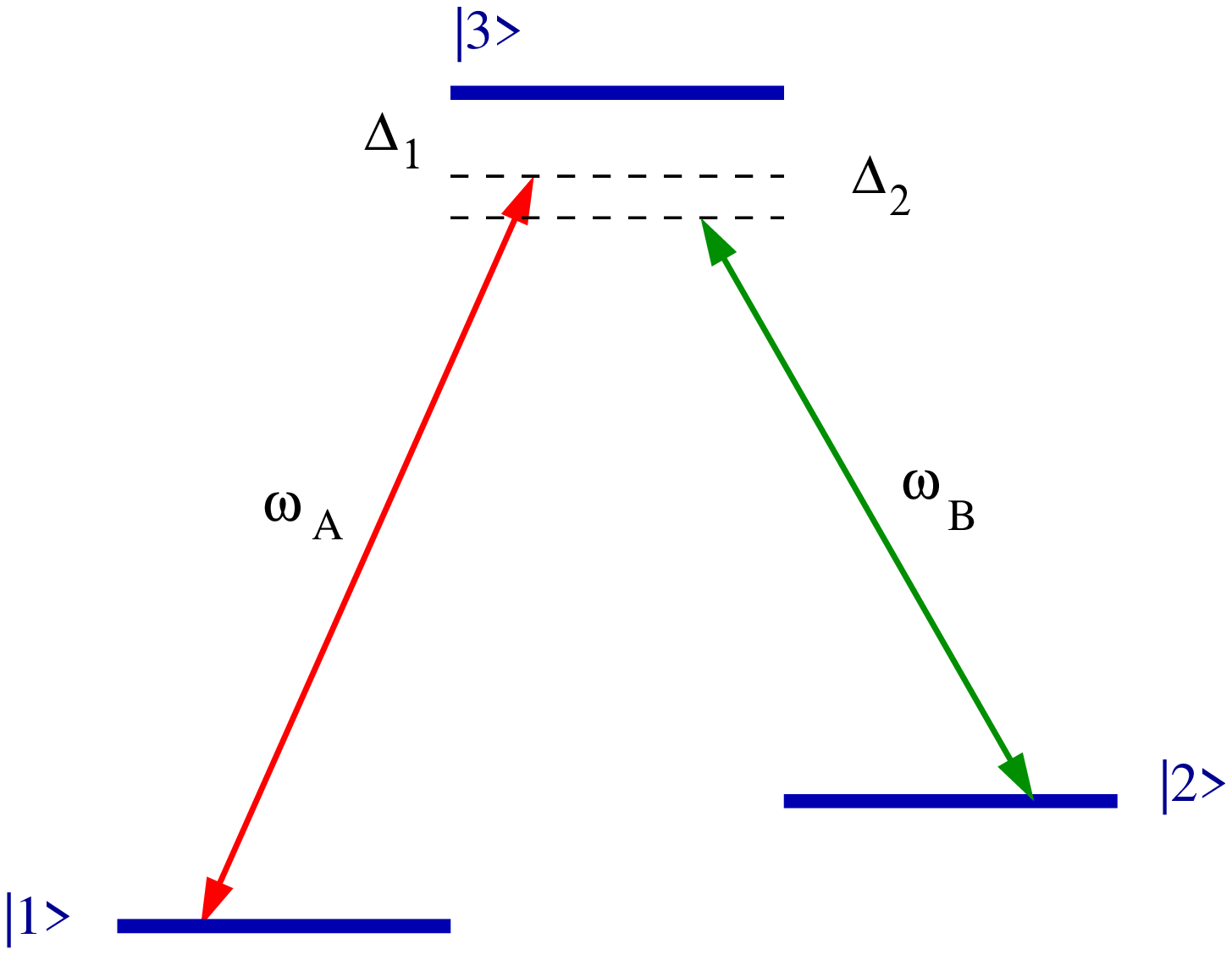}
\caption{Atomic energy levels in a $\Lambda$-system. The laser frequencies
$\omega_A$ and $\omega_B$ are detuned from the atomic energy splittings
$\omega_{31}$ and $\omega_{32}$ by $\Delta_1$ and $\Delta_2$ respectively.
The differential detuning is $\d = \D_2 - \D_1$.}
}

Here, we have three atomic energy levels together with two lasers with 
frequencies $\w_A$ and $\w_B$ tuned nearly, but not exactly, 
to the frequency differences 
$\w_{31} = \w_3 - \w_1$ and $\w_{32} = \w_3 - \w_2$ of the levels.
One of these lasers will be the `coupling' (or `pump') field while the
other will be a `probe' beam whose propagation through the coupled
laser-atom system we wish to study. Experimentally, the atomic levels are
typically realised by suitable hyperfine levels in ultracold gases of
alkali metal atoms, Na, Rb, Cs.

The principal application of this set-up is to electromagnetically-induced
transparency (EIT). For a review, see \cite{Marangos}.
This allows the well-publicised phenomenon of `slow light'. 
In this case, laser B is a strong, coupling field and A is the probe. 
This gives rise to a double absorption line with an intermediate region
where $\Ima n(\w) = 0$ (transparency) and the refractive index is linear
and rapidly rising (normal dispersion), giving rise to an extremely small 
group velocity for the probe light. This is the physics behind the
slow light experiments. These have achieved group velocities as low as
a few ms$^{-1}$ in an ultracold gas of Na atoms \cite{Hau}.
Subsequent experiments \cite{Liu,Phillips} 
have even demonstrated the possibility of stopping and reconstructing a 
light pulse in a $\L$-system in either ultracold Na or warm Rb vapour cells.

For our purposes, we are more concerned with the opposite case where
A is the coupling (or Raman pump) field while B is the probe laser.
Here, we find a single {\it Raman gain line}, i.e. 
$\Ima n(\w) < 0$. This demonstrates how it is possible to achieve
the desired condition $v_{\rm ph}(\infty) < v_{\rm ph}(0)$.
Although not directly relevant to our discussion here, this also allows
the remarkable phenomenon of a negative group velocity. To achieve this,
we introduce two coupling fields so that we have a gain doublet with
an intermediate region characterised by virtual transparency with 
a strong linear anomalous dispersion. If the slope of $\Rea n(\w)$ is 
sufficiently large and negative, this gives a negative group velocity --
a light pulse traversing a cell with this coupled laser-atom system is 
advanced sufficiently that its peak leaves the cell before entering it.
Although certainly counter-intuitive, this astonishing effect is not
at all incompatible with causality and has been demonstrated experimentally
by Wang {\it et al.} \cite{Wang,Dogariu,Wangvenice} 
in $\L$-systems in warm atomic Cs gas. 

\vskip0.2cm
We now present a unified analysis of $\L$-systems which allows us to describe
both the EIT and Raman gain scenarios in a common framework. EIT is 
obtained from our general formulae in one limit, while the complementary
limit describes Raman gain lines.

It is most convenient to use a density matrix formulism. Writing the
wave function for the 3-state $\L$-system shown in Fig.~6 as
\begin{equation}
|\psi(t)\rangle ~=~ c_1(t) e^{-i\w_1 t}|1\rangle 
+ c_2(t) e^{-i\w_2 t}|2\rangle + c_3(t) e^{-i\w_3 t}|3\rangle
\label{eq:fa}
\end{equation}
we define density matrix elements as $\r_{ij} = c_i c_j^*$.
The diagonal elements $\r_{11}, \r_{22}, \r_{33}$ are referred to as the
`populations' while the off-diagonal elements $\r_{31}, \r_{32}, \r_{21}$
are the `coherences'. It is also useful to introduce the notation
$\hat \s_{ij} = |i\rangle\langle j|$.

The laser-atom interaction is described by the 
potential $V = e {\underline r} . {\underline E} \cos\w t$, where $\w$
is the laser frequency, with matrix elements 
\begin{equation}
V_{ij} ~=~ e \m_{ij}|{\underline E}|\cos\w t ~=~ 
\W_{ij}\cos\w t 
\label{eq:fb}
\end{equation}
which defines the Rabi frequencies $\W_{ij}$ in terms of the dipole
matrix elements $\m_{ij}$.
The polarisation of this laser-atom system is given in terms of the density
matrix elements by
\begin{equation}
P ~=~ N \bigl(\m_{13} \r_{31} e^{-i\w_{31}t} + 
\m_{23}\r_{32} e^{-i\w_{32}t} + ~c.c.~\bigr)
\label{eq:fc}
\end{equation}
where $N$ is the atomic number density, since the diagonal elements do
not contribute by the dipole selection rule. Abbreviating $\W_{31} = \W_A$
and $\W_{32} = \W_B$ for convenience, and anticipating the solution of
the Schrodinger equation for the time dependence of the density matrix
to set
\begin{equation}
\r_{31} ~=~ \W_A e^{i\D_1 t}\hat\r_{31} ~~~~~~~~~~~~~~~~
\r_{32} ~=~ \W_B e^{i\D_2 t}\hat\r_{32}
\label{eq:fd}
\end{equation}
we have
\begin{equation}
P ~=~ N\bigl(\m_{13} \W_A e^{-i\w_A t} \hat\r_{31} +
\m_{23}\W_B e^{-i\w_B t}\hat\r_{32} ~+~c.c.~\bigr)
\label{eq:fe}
\end{equation}
Recalling that in general $P = \chi \e_0 |\underline{E}| \cos\w t$,
we find the dielectric susceptibilities for a probe laser A in the 
coupled laser B--atom system, and vice-versa, are
\begin{equation}
\chi_A ~=~ {2N\over\e_0} |\m_{13}|^2 \hat\r_{31} ~~~~~~~~~~~~~~~~
\chi_B ~=~ {2N\over\e_0} |\m_{23}|^2 \hat\r_{32} 
\label{eq:ff}
\end{equation}
This is related to the refractive index by $n(\w) = 1 + \chi/2$, so we 
eventually find the following essential relations between the refractive 
indices for lasers A and B and the reduced density matrix elements 
$\hat\r_{ij}$:
\begin{equation}
n_A(\w_A) ~=~ 1 + {N\over\e_0} |\m_{13}|^2 \hat\r_{31} ~~~~~~~~~~~~~~~~
n_B(\w_B) ~=~ 1 + {N\over\e_0} |\m_{23}|^2 \hat\r_{32}
\label{eq:fg}
\end{equation}

The next step is to calculate the density matrix elements. They satisfy
the Schrodinger equation 
\begin{equation}
{d\r\over dt} ~=~ [H_I,\r]
\label{eq:fh}
\end{equation}
where the laser-atom interaction (\ref{eq:fb}) is (implementing the 
standard `rotating-wave approximation' \cite{Foot})
\begin{equation}
H_I ~=~ -{1\over2} \bigl( \W_A e^{i\D_1 t} \hat\s_{31} +
\W_B e^{i\D_2 t} \hat\s_{32} ~+~ c.c.~\bigr)
\label{eq:fi}
\end{equation}
The important equations are those for the coherences. With the initial
condition $\r_{11} = 1, ~\r_{22} = \r_{33} = 0$ corresponding to a
population initially in state $|1\rangle$, we find
\begin{eqnarray}
{d\r_{31}\over dt} ~=~ {i\over2} \W_A e^{i\D_1 t} + 
{i\over2} \W_B e^{i\D_2 t} \r_{21} - {1\over2}\c_{31}\r_{31} 
~~~~~~~~~~\nonumber\\
{}
\nonumber\\
{d\r_{32}\over dt} ~=~  
{i\over2} \W_B e^{i\D_1 t} \r_{21}^* - {1\over2}\c_{32}\r_{32} 
~~~~~~~~~~~~~~~~~~~~~~~~~~\nonumber\\
{}
\nonumber\\
{d\r_{21}\over dt} ~=~ -{i\over2} \W_A e^{i\D_1 t} \r_{32}^* +  
{i\over2} \W_B^* e^{-i\D_2 t} \r_{31} - {1\over2}\c_{21}\r_{21}
\label{eq:fj}
\end{eqnarray}
where we have included the total widths $\c_{ij}$ describing the
spontaneous radiative decay of state $|3\rangle$ and dephasing \cite{Marangos}.
It is straightforward to see that the time-dependence implied by these
equations is consistent with eq.(\ref{eq:fd}). For convenience, we also
set $\r_{21} = \exp(i\d t) \hat\r_{21}$, where $\d = \D_1 - \D_2$.
Then, with notation $\tilde \D_1 = \D_1 - {i\over2}\c_{31}$ etc., we obtain
the following set of algebraic equations for the reduced matrix elements
$\hat\r_{ij}$:
\begin{eqnarray}
&2\tilde\D_1 \hat\r_{31} ~&=~ 1 + {\W_B\over\W_A} \hat\r_{21}
\nonumber\\
&2\tilde\D_2 \hat\r_{32} ~&=~ {\W_A\over\W_B} \hat\r_{21}^*
\nonumber\\
&2\tilde\d ~\hat\r_{21} ~&=~ - \W_A \W_B^* \hat\r_{32}^* +
\W_B^* \W_A \hat\r_{31}
\label{eq:fk}
\end{eqnarray}
Solving these by elimination of $\hat\r_{21}$, we find the master formulae
relevant for a general $\L$-system:
\begin{eqnarray}
&\hat\r_{31} ~&=~ {1\over2\tilde\D_1} 
\biggl(1 + {|\W_A|^2\over4\tilde\D_2^* \tilde\d}\biggr)~
\biggl[1 + {|\W_A|^2\over4\tilde\D_2^* \tilde\d} -
{|\W_B|^2\over4\tilde\D_1^* \tilde\d}\biggr]^{-1}
\nonumber\\
&{}&{}
\nonumber\\
&\hat\r_{32} ~&=~ {|\W_A|^2\over 8 \tilde\D_1^* \tilde\D_2 \tilde\d^*}
\biggl[1 + {|\W_A|^2\over4\tilde\D_2^* \tilde\d} -
{|\W_B|^2\over4\tilde\D_1^* \tilde\d}\biggr]^{-1}
\label{eq:fl}
\end{eqnarray}
At this point, we specialise to the two cases of interest -- EIT and
slow light, and Raman gain lines.

\subsection{EIT and slow light}

The standard EIT scenario is where $B$ is the coupling laser while
A is a weak probe. In this case, we are interested in the refractive index
$n_A(\w_A)$ for propagation of the probe laser light through the 
coupled laser B--atom system. This is given by the reduced density matrix
element $\hat\r_{31}$. Since the probe laser is weak, we assume
$|\W_A|^2/\D_2\d \ll 1$ and drop the corresponding term in eq.(\ref{eq:fl}).
We also assume that $\c_{21} \simeq 0$. This gives
\begin{equation}
\hat\r_{31} ~=~ {1\over2} ~
{1\over \D_1 - {|\W_B|^2\over4\d} - {i\over2}\c_{31}}
\label{eq:fm}
\end{equation}
In terms of the refractive index, this is:
\begin{eqnarray}
&\Rea n_A(\w_A) ~&=~ 1 ~+~ {N\over2\e_0} |\m_{13}|^2 ~
{\Bigl(\D_1 \d - {|\W_B|^2\over4}\Bigr)\d \over
\Bigl(\D_1 \d - {|\W_B|^2\over4}\Bigr)^2 + 
{1\over4} \c_{31}^2 \d^2 }
\nonumber\\
&{}&{}
\nonumber\\
&\Ima n_A(\w_A) ~&=~  {N\over4\e_0} |\m_{13}|^2 ~
{\c_{31}\d^2 \over
\Bigl(\D_1 \d - {|\W_B|^2\over4}\Bigr)^2 + 
{1\over4} \c_{31}^2 \d^2 }
\label{eq:fn}
\end{eqnarray}

The dependence on the probe frequency $\w_A = \w_{31} - \D_1$, 
for fixed $\w_B$, is sketched in Fig.~7.
The main feature in $\Ima n_A(\w_A)$ is a double absorption line with
an intermediate region around $\w_A \simeq \w_B + \w_{21} ~~(\d = 0)$ where 
$\Ima n_A(\w_A) \simeq 0$. At $\d = 0$, the imaginary part vanishes
so there is no absorption. This is electromagnetically-induced transparency. 

\FIGURE
{\epsfxsize=7cm\epsfbox{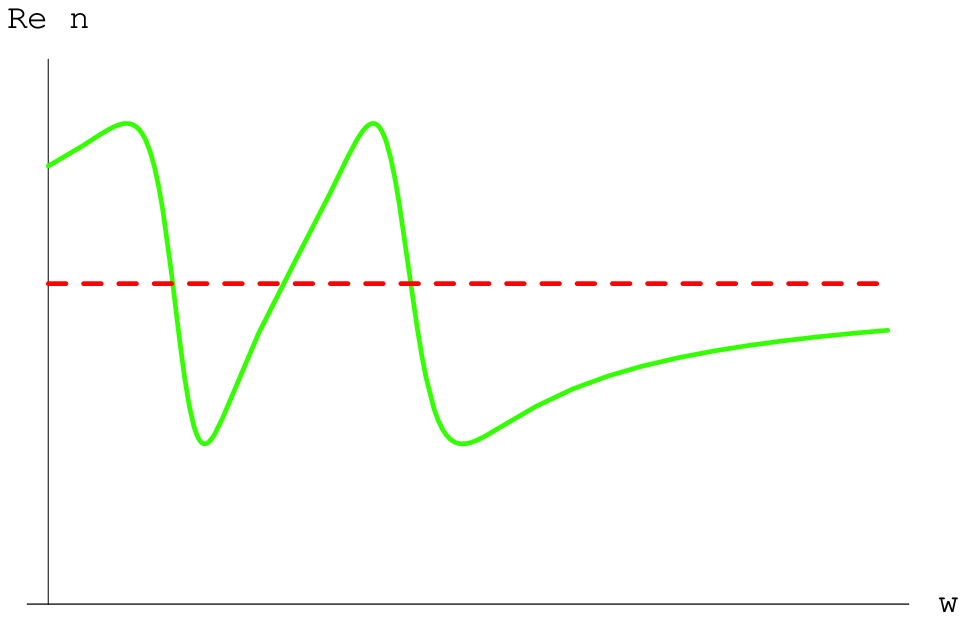}\hskip1cm
\epsfxsize=7cm\epsfbox{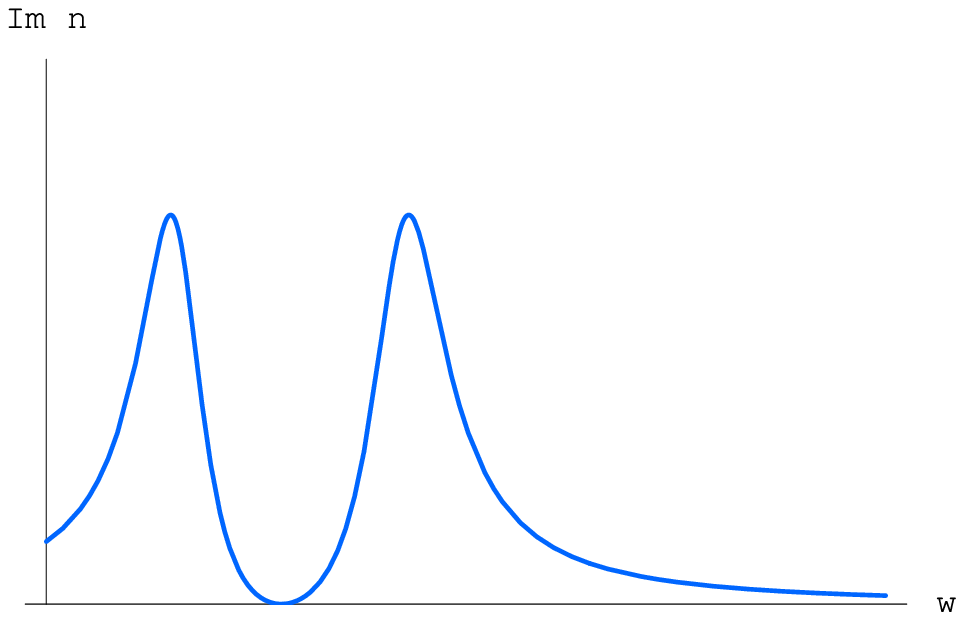}
\caption{The real (left figure) and imaginary (right figure) parts of the
refractive index in a $\L$-system exhibiting electromagnetically-induced
transparency (EIT).}
}

However, although there is near complete transparency in this frequency 
region, there is significant dispersion -- the refractive index 
$\Rea n_A(\w_A)$ rises steeply (normal dispersion) and nearly linearly.
Recalling that the group velocity is given by 
\begin{equation}
v_{\rm gp}(\w) ~=~ \Bigl( n + \w {dn\over d\w}\Bigr)^{-1}
\label{eq:fo}
\end{equation}
a large positive slope $dn/d\w$ corresponds to a small group velocity.
Linearity implies negligible pulse distortion. This is the phenomenon
of `slow light', which has allowed group velocities as low as 17 ms$^{-1}$
to be achieved in ultracold Na atom cells \cite{Hau}.

\subsection{Raman gain lines and gain-assisted anomalous dispersion}

To describe the Raman gain scenario, we reverse the roles of the lasers so 
that now A corresponds to the coupling field and B is the probe.
This time, therefore, we are interested in the refractive index $n_B(\w_B)$
given by the reduced coherence $\hat\r_{32}$ in eq.(\ref{eq:fl}).
In the relevant experiment \cite{Wang,Dogariu}, we are concerned with
the two-photon Raman transition $|1\rangle \rta |2\rangle$ via the
intermediate state $\sim |3\rangle$, so laser A is interpreted as the 
Raman pump and B the Raman probe field. 

The relevant approximation here is that (while $|\W_A| \gg| \W_B|$)
both fields are weak in the sense $|\W_A|^2/\D_2\d \ll 1$ and
$|\W_B|^2/\D_1\d \ll 1$, while for such Raman experiments we can neglect the
width of the state $|3\rangle$ so that $\c_{31} \simeq \c_{32} \simeq 0$.
We also assume the common detuning $\bar \D = (\D_1 + \D_2)/2$ is much
greater than the differential detuning $\d = \D_1 - \D_2$.

With these simplifications, eq.(\ref{eq:fl}) reduces to 
\begin{equation}
\hat\r_{32} ~=~ {|\W_A|^2\over 8\bar\D^2}~{1\over \d + {i\over2}\c_{21}}
\label{eq:fp}
\end{equation}
The corresponding refractive index is
\begin{eqnarray}
&\Rea n_B(\w_B) ~&=~ 1 ~+~ {N\over8\e_0} |\m_{23}|^2 {|\W_A|^2\over\bar\D^2}~
{\d \over \d^2 + {1\over4}\c_{21}^2}
\nonumber\\ 
&{}&{}
\nonumber\\
&\Ima n_B(\w_B) ~&=~ - {N\over16\e_0} |\m_{23}|^2 {|\W_A|^2\over\bar\D^2}~
{\c_{21} \over \d^2 + {1\over4}\c_{21}^2}
\label{eq:fq}
\end{eqnarray}
The refractive index is sketched as a function of $\w_B = \w_{32} - \D_2$,
for fixed $\w_A$, in Fig.~8.
\FIGURE
{\epsfxsize=7cm\epsfbox{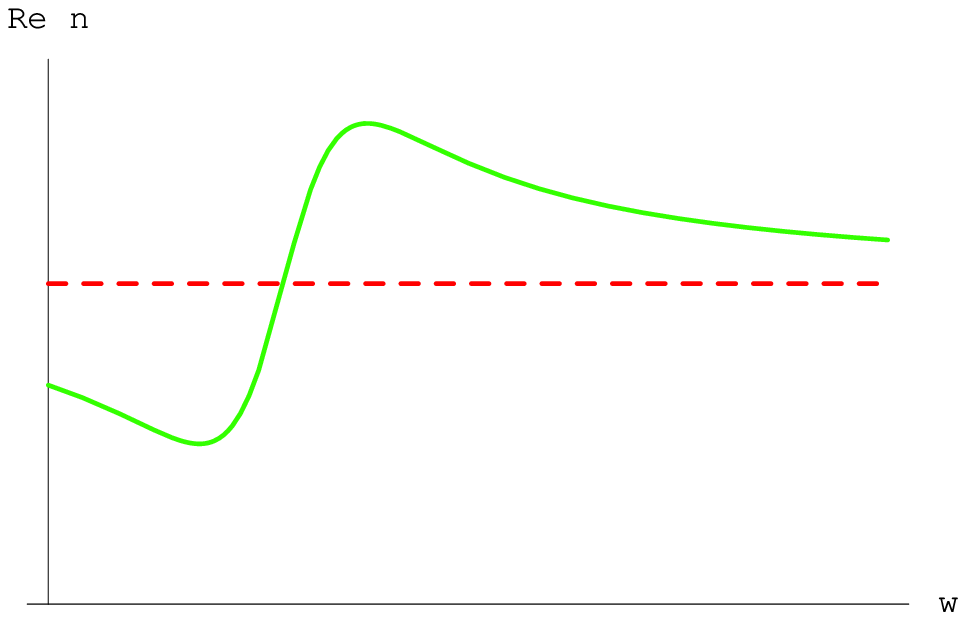}\hskip1cm
\epsfxsize=7cm\epsfbox{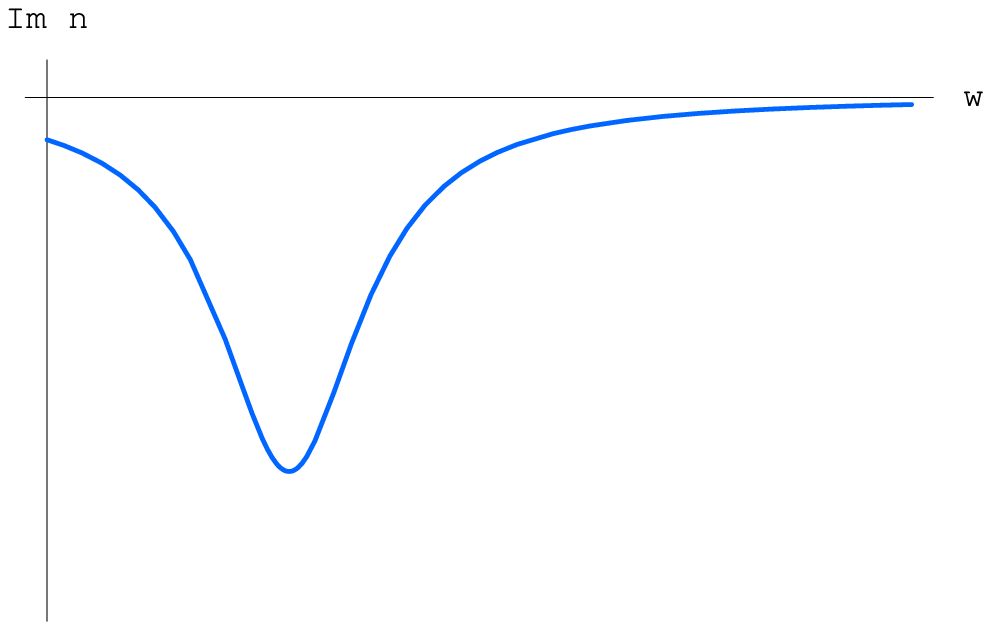}
\caption{The real (left) and imaginary (right) parts of the refractive
index for the single Raman gain line given by eq.(\ref{eq:fq}). This
illustrates how it is possible to have $v_{\rm ph}(\infty) < v_{\rm ph}(0)$
in a system with $\Ima n(\w) < 0$.}
}

Here, the physics is quite different. This time we have $\Ima n_B(\w_B) < 0$,
indicating {\it gain} rather than absorption. The single Raman gain line
is centred on $\w_B \simeq \w_A - \w_{21}~~(\d =0)$. For low frequencies,
the refractive index $\Rea n_B(\w_B \simeq 0)$ is less than 1, corresponding
to a superluminal phase velocity $v_{\rm ph}(\w_B \simeq 0) > 1$.
For large $\w_B$, the phase velocity tends to 1 as usual,
$v_{\rm ph}(\infty) = 1$. This behaviour of the refractive index and
phase velocity is therefore compatible with the KK dispersion relation
with $\Ima n_B(\w_B) < 0$.

The Raman gain line therefore exhibits precisely the behaviour we speculate
is necessary to resolve the potential paradox associated with the
superluminal phase velocity in QED in a gravitational field.
The essential feature is that the coupled laser-atom $\L$-system in this
case exhibits gain rather than absorption, i.e. $\Ima n(\w) < 0$
rather than the more familiar $\Ima n(\w) > 0$.
Unlike the situation with absorption, where photons are scattered out of
the probe beam, in the case of gain the intensity of the probe beam is
increased as it passes through the laser-atom medium. This amplification of
the electric field of the probe laser is due to the additional photons
in the probe beam coming from the second stage of the Raman transition. 
It is maximised when the differential detuning $\d \simeq 0$.

The crucial point for our considerations is that it is possible to have
$v_{\rm ph}(\infty) < v_{\rm ph}(0)$ provided $\Ima n(\w) < 0$, which
means that the light must gain in intensity due to its passage through the 
medium. In the next section, we discuss whether this physical mechanism
could be realised in the very different scenario of light propagation
in gravitational fields.

\vskip0.2cm
Before leaving this section, however, we briefly mention an intriguing
experimental application of the Raman gain scenario. 
This involves a system with {\it two} Raman pump fields
with slightly different frequencies. This produces a {\it gain doublet},
with refractive index as sketched in Fig.~9.

\FIGURE
{\epsfxsize=7cm\epsfbox{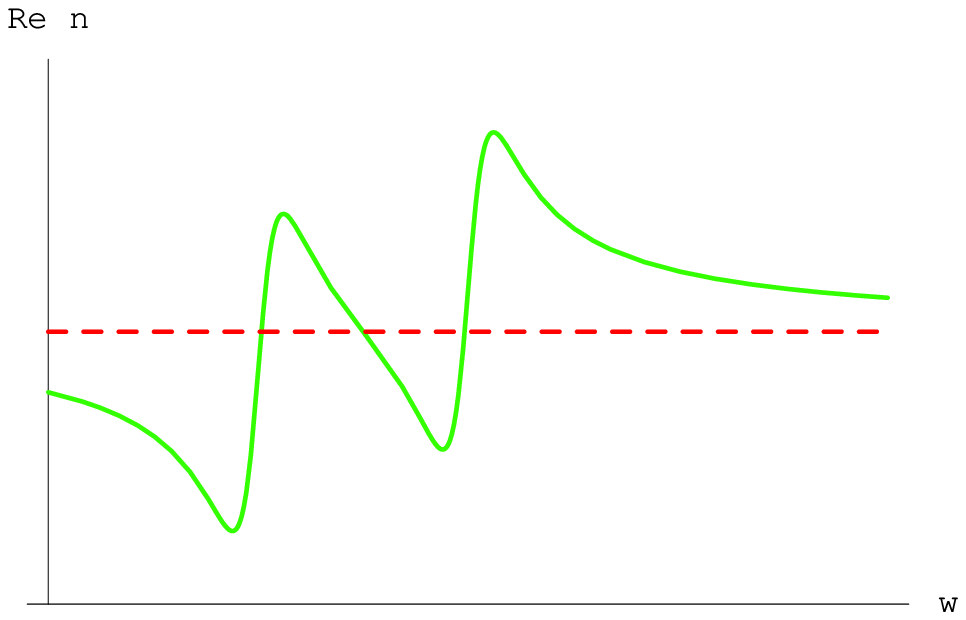}\hskip1cm
\epsfxsize=7cm\epsfbox{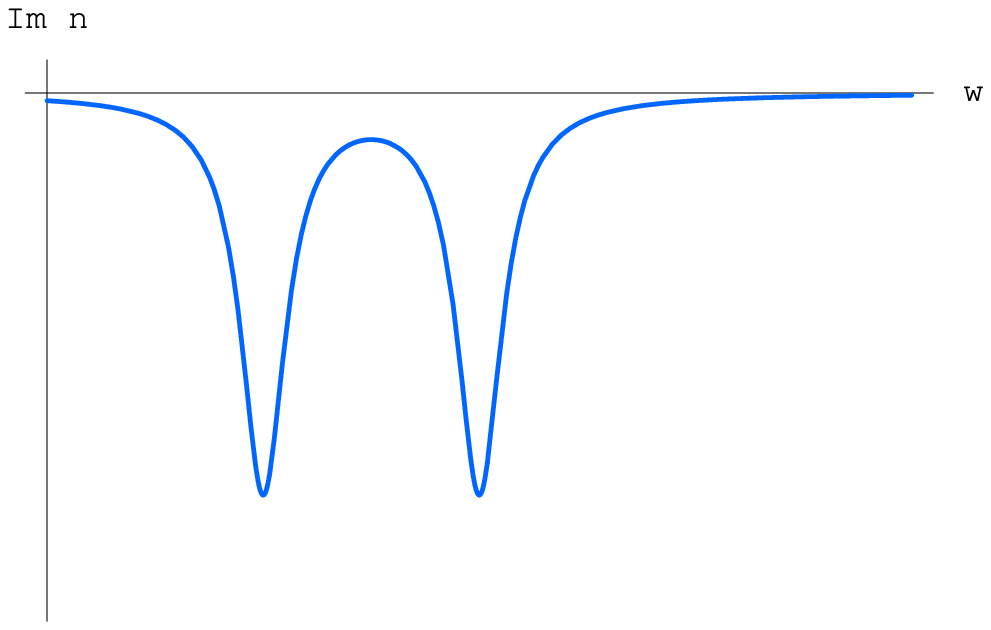}
\caption{The real (left) and imaginary (right) parts of the refractive
index for the Raman gain doublet, exhibiting the phenomenon of gain-assisted
anomalous dispersion and negative group velocity.}
}

The region between the two gain lines has $\Ima n(\w) \simeq 0$, so there
is virtually no gain, or absorption, and once again the system is
transparent to the probe light. The refractive index in this region is
steeply falling, and can be made approximately linear as shown -- we 
therefore have a frequency range exhibiting transparency with linear
anomalous dispersion. From eq.(\ref{eq:fo}), we see that the group velocity
in this region may be made extremely large and even, if the slope
$dn/d\w$ is sufficiently large, we may find $v_{\rm gp}$ {\it negative}.

Remarkably, the phenomenon of a negative group velocity with negligible
pulse distortion has been realised experimentally 
\cite{Wang,Dogariu,Wangvenice}. The set-up for these experiments uses a 
$\L$-system where the energy levels shown in Fig.~6 are  
hyperfine levels of Cs atoms with $|1\rangle = 6S_{1/2}|F=4, m=-4\rangle$,~
$|2\rangle = 6S_{1/2}|F=4, m=-2\rangle$ and 
$|3\rangle = 6P_{3/2}|F=4, m=-3\rangle$
with the probe laser close to resonance with the Cs $D_2$ line. A laser 
pulse was passed through the Cs vapour cell with a group velocity of 
$- 0.003c$, so that it exits the cell {\it before} entering it.
For more details of this intriguing experiment, 
see refs.\cite{Wang,Dogariu,Wangvenice}.

\section{Superluminality and UV completion}

In this paper, we have analysed the propagation of light and the dispersion
relation for the refractive index in three distinct quantum systems.
For QED in a background electromagnetic field, we have used a recent
evaluation of the one-loop vacuum polarisation to plot both the real and
imaginary parts of $n(\w)$ over the full frequency range. Expressing
the vacuum polarisation in the Newman-Penrose formalism allowed an 
elegant identification of the role of the null energy projection
$T_{\m\n}k^\m k^\n$, clarifying its relation to the possible occurrence
of superluminal propagation. For QED in a gravitational field, only the
low-frequency range of the refractive index is currently well understood.
We reviewed and extended some of our earlier work in this field,
demonstrating the occurrence of a superluminal low-frequency phase velocity,
$v_{\rm ph}(0) > 1$, for certain spacetimes. This was illustrated with 
examples of both Ricci-flat and Weyl-flat spacetimes, the former 
exhibiting gravitational birefringence as well as superluminality.
Finally, we gave a unified treatment of propagation in coupled laser-atom
$\L$-systems exhibiting either electromagnetically-induced transparency with
`slow light' or Raman gain lines, where gain-assisted anomalous dispersion
permitted the remarkable and experimentally observed phenomenon of a
negative group velocity.

The purpose of drawing these various examples together here is to shed light 
on the proposal that the fundamental axioms of local quantum field theory or
string theory imply constraints on the couplings of the corresponding IR 
effective field theory. That is, does the requirement that an effective field 
theory should have a consistent UV completion necessarily imply
positivity constraints on the couplings of the leading irrelevant operators,
with implications for the phenomenology of the low-energy theory?  

In the case of QED in flat spacetime, this proposal works beautifully.
As we explained, causality itself only requires the {\it UV limit} of the
phase velocity to be subluminal, $v_{\rm ph}(\infty) < 1$. But provided
the imaginary part of the refractive index is positive, which we
confirmed by an explicit evaluation showing $\Ima n(\w)$ has a single
absorption line in a background electromagnetic field, the KK dispersion
relation ensures $v_{\rm ph}(0) < v_{\rm ph}(\infty)$. So causality does
indeed require $v_{\rm ph}(0) < 1$, which implies positivity constraints
on the leading irrelevant operators of the low-energy effective theory
for QED, viz.~the Euler-Heisenberg effective action. Of course, explicit
perturbative calculations confirm that these are satisfied.

QED in curved spacetime, on the other hand, presents a serious problem.
Here we find examples where the low-energy effective theory has a 
superluminal phase velocity, $v_{\rm ph}(0) > 1$. At first sight, this 
appears to be incompatible with causality. We identified three ways in
which causality could be maintained: first, that an extension of the 
notion of `stable causality' may permit $v_{\rm ph}(\infty) > 1$ without
causal violations; second, that the KK dispersion relation is not valid
in its standard form for QFT in curved spacetime; and third, that at least
in gravitational theories, $\Ima n(\w)$ (and by extension $\Ima \MM(s,0)$)
may be {\it negative}. To explore the plausibility of this last option,
we showed how a negative $\Ima n(\w)$ arises in Raman $\L$-systems,
where it is associated with gain rather than absorption. We have not found
a similar example of $\Ima n(\w) < 0$ in a non-gravitational quantum field
theory and speculate that its occurrence for QED in curved spacetime may
be related to the ability of gravitational interactions to focus geodesics
and amplify light waves.

We do not yet know with certainty which of these is the physically realised
option. However, each of them would invalidate the use of $v_{\rm ph}(0) < 1$
(or related analyticity constraints on scattering amplitudes $\MM(s,t)$)
as a criterion constraining the couplings of the IR effective field theory,
since we must allow the possibility that the UV completion may involve 
gravity. This would appear to undermine the use of the positivity 
constraints as restrictions on low-energy phenomenology, though with better
understanding they may still provide important information on the nature
of the UV theory. In any case, understanding how to accommodate a 
superluminal low-frequency phase velocity with causality in quantum field 
theory is now an urgent problem whose resolution will surely shed light on 
the quantum theory of gravity.

\vfill\eject

\appendix

\section{Appendix:~Electrodynamics in Newman-Penrose formalism}

We collect here some formulae involving the Newman-Penrose formalism
which are used in the main text. A useful review is contained in  
ref.\cite{Chandra}. The essential 
feature is the introduction of a null tetrad with vierbeins $e_A^\m$ 
where the index `$A$' refers to four null vectors $\ell^\m, n^\m, m^\m, 
\bar m^\m$. These are chosen so that their only non-vanishing scalar 
products are $\ell.n = 1$ and $m.\bar m = -1$.\footnote{In flat spacetime,
a standard representation of these vectors would be 
$\ell^\m = {1\over\sqrt2} (1,0,0,1),~ n^\m = {1\over\sqrt2}(1,0,0,-1),~ 
m^\m = {1\over\sqrt2} (0,1,i,0),~ \bar m^\m = {1\over\sqrt2} (0,1,-i,0)$.
This is the origin of the factor $\sqrt2$ when we express the wave-vector
as $k^\m = \sqrt2 \w\ell^\m$, identifying $\w$ as the frequency 
(see section 2). We use a metric $g_{\m\n}$ with signature $(+,-,-,-)$.} 
In this basis, the metric is therefore
\begin{equation}
\eta_{AB} ~=~ e_A^\m e_B^\n g_{\m\n} ~=~ 
\left(\matrix{0&1&0&0\cr  1&0&0&0\cr  0&0&0&-1\cr  0&0&-1&0\cr}\right)
\label{eq:ha}  
\end{equation}
It follows that
\begin{equation}
g_{\m\n} ~=~ (\ell_\m n_\n) - (m_\m \bar m_\n)  
\label{eq:hb}
\end{equation}
with notation $(ab) = ab + ba$. We also use $[ab] = ab - ba$.
It follows that the trace in the null basis is~ 
${\tr} M_{AB} = \eta^{BA}M_{AB} = M_{(\ell n)} - M_{(m \bar m)}$.

To describe the electromagnetic field strengths, we introduce three 
complex scalars $\phi_0, \phi_1, \phi_2$ defined as
\begin{eqnarray}
&\phi_0 ~=~ F_{\m\n} \ell^\m m^\n ~~~~~~~~~~~~~~~\nonumber \\
&\phi_1 ~=~ {1\over2} F_{\m\n} (\ell^\m n^\n + \bar m^\m m^\n) \nonumber \\
&\phi_2 ~=~ F_{\m\n} \bar m^\m n^\n ~~~~~~~~~~~~~~
\label{eq:hc}  
\end{eqnarray}
Inverting this, we recover the formula for $F_{\m\n}$ quoted in 
eq.(\ref{eq:bu}), viz.
\begin{equation}
F^{\m\n} ~=~ -(\phi_1 + \phi_1^*) [\ell^\m n^\n] + 
(\phi_1 - \phi_1^*)[m^\m \bar m^\n] +\phi_2 [\ell^\m m^\n] 
+ \phi_2^* [\ell^\m \bar m^\n] - \phi_0^*[n^\m m^\n] - 
\phi_0 [n^\m \bar m^\n]  
\label{eq:hd}  
\end{equation}
In terms of the vierbein basis, this is
\begin{equation}
F_{AB} ~=~ \left(\matrix{0 & (\phi_1 + \phi_1^*) & \phi_0 & \phi_0^* \cr
-(\phi_1 + \phi_1^*) & 0 & -\phi_2^* & -\phi_2 \cr
-\phi_0 & \phi_2^* & 0 & -(\phi_1 - \phi_1^*) \cr
-\phi_0^* & \phi_2 & (\phi_1 - \phi_1^*) & 0 \cr}\right)
\label{eq:he}   
\end{equation}

It is useful to write explicit expressions for the $\phi_i$ in terms of
the usual field strengths ${\bf E}$ and ${\bf B}$ in flat spacetime.
Defining 
\begin{equation}
F_{\m\n} ~=~ \left(\matrix{0 & E_1 & E_2 & E_3 \cr
-E_1 & 0 & B_3 & -B_2 \cr -E_2 & -B_3 & 0 & B_1 \cr
-E_3 & B_2 & - B_1 & 0\cr}\right)  
\label{eq:hf}
\end{equation}
we find
\begin{eqnarray}
E_1 ~=~ \Rea (\phi_0 - \phi_2) ~~~~~~~~~~B_1 ~=~ - \Ima (\phi_0 - \phi_2) 
\nonumber \\
E_2 ~=~ \Ima (\phi_0 + \phi_2) ~~~~~~~~~~B_2 ~=~ \Rea (\phi_0 + \phi_2)~~~~
\nonumber \\
E_3 ~=~ -2 \Rea \phi_1 ~~~~~~~~~~~~~~B_3 ~=~ 2 \Ima \phi_1 ~~~~~~~~~~~
\label{eq:hg} \\  
\end{eqnarray}
and the inverse relations
\begin{eqnarray}
\phi_0 ~=~ {1\over2}\Bigl( (E_1 + B_2) + i(E_2 - B_1)\Bigr)~~~~ \nonumber \\
\phi_1 ~=~ -{1\over2} (E_3 - i B_3) ~~~~~~~~~~~~~~~~~~~~~ \nonumber \\
\phi_2 ~=~ {1\over2}\Bigl( -(E_1-B_2) + i (E_2 + B_1)\Bigr) ~~
\label{eq:hh}  
\end{eqnarray}

The corresponding results for the dual field strength tensor 
$\tilde F_{\m\n} = {1\over2} \e_{\m\n\l\r} F^{\l\r}$ are
\begin{equation}
\tilde F^{\m\n} ~=~ -i(\phi_1 - \phi_1^*) [\ell^\m n^\n] + 
i(\phi_1 + \phi_1^*)[m^\m \bar m^\n] +i\phi_2 [\ell^\m m^\n] 
-i \phi_2^* [\ell^\m \bar m^\n] + i \phi_0^*[n^\m m^\n] - 
i \phi_0 [n^\m \bar m^\n]   
\label{eq:haa}
\end{equation}
that is,
\begin{equation}
\tilde F_{AB} ~=~ \left(\matrix{0 & i(\phi_1 - \phi_1^*) 
& i\phi_0 & -i\phi_0^* \cr
-i(\phi_1 - \phi_1^*) & 0 & i\phi_2^* & -i\phi_2 \cr
-i\phi_0 & -i\phi_2^* & 0 & -(\phi_1 + \phi_1^*) \cr
i\phi_0^* & i\phi_2 & i(\phi_1 + \phi_1^*) & 0 \cr}\right)  
\label{eq:hbb}
\end{equation}

\vskip0.3cm
We also need expressions for the NP components of the squares of
the field strengths $(F^2)_{\m\n}$, $(F\tilde F)_{\m\n}$ and
$(\tilde F \tilde F)_{\m\n}$. In matrix notation, we find
\begin{equation}
(F^2)_{AB} ~=~  \left(\matrix{2\phi_0 \phi_0^* 
&(F^2)_{\ell n} 
& 2\phi_0 \phi_1^*  & 2 \phi_1 \phi_0^* \cr
(F^2)_{n\ell}
& 2\phi_2 \phi_2^* & 2\phi_1 \phi_2^* & 2\phi_2 \phi_1^* \cr
2\phi_0 \phi_1^* & 2\phi_1 \phi_2^* & 2\phi_0 \phi_2^* 
& (F^2)_{m \bar m} \cr
2\phi_1 \phi_0^* & 2\phi_2 \phi_1^* 
& (F^2)_{\bar m  m}
& 2\phi_2 \phi_0^* \cr}\right)  
\label{eq:hcc}
\end{equation}
where $(F^2)_{\ell n} = (F^2)_{n\ell}= 
(\phi_1 + \phi_1^*)^2 - \phi_0 \phi_2 - \phi_0^* \phi_2^*$ and
$(F^2)_{m\bar m} =  (F^2)_{\bar m m} = 
-(\phi_1 - \phi_1^*)^2 + \phi_0 \phi_2 + \phi_0^* \phi_2^*$.
Also,
\begin{equation}
(F\tilde F)_{AB} ~=~ i \left(\matrix{0 & (F\tilde F)_{\ell n} &0&0 \cr
(F\tilde F)_{n\ell} &0&0&0\cr
0&0&0& (F\tilde F)_{m\bar m}\cr
0&0 & (F\tilde F)_{\bar m m} &0 \cr}\right)  
\label{eq:hdd}
\end{equation}
where $(F\tilde F)_{\ell n}= (F\tilde F)_{n \ell} =
(F\tilde F)_{m \bar m} = (F\tilde F)_{\bar m m} =
(\phi_1 + \phi_1^*)(\phi_1 - \phi_1^*) - \phi_0 \phi_2
+ \phi_0^* \phi_2^*$.
Finally,
\begin{equation}
(\tilde F \tilde F)_{AB} ~=~ \left(\matrix{2\phi_0 \phi_0^*
& (\tilde F \tilde F)_{\ell n} 
& 2 \phi_0 \phi_1^* & 2 \phi_1 \phi_0^* \cr
(\tilde F \tilde F)_{n \ell} & 2 \phi_2 \phi_2^* & 2 \phi_1 \phi_2^*
& 2 \phi_2 \phi_1^* \cr
2 \phi_0 \phi_1^* & 2 \phi_1 \phi_2^* & 2 \phi_0 \phi_2^*
& (\tilde F \tilde F)_{m \bar m} \cr
2\phi_1 \phi_0^* & 2 \phi_2 \phi_1^* 
& (\tilde F \tilde F)_{\bar m m} & 2\phi_2 \phi_0^* \cr}\right)
\label{eq:hee}  
\end{equation}
where here  $(\tilde F \tilde F)_{\ell n} =  (\tilde F \tilde F)_{n\ell}
= -(\phi_1 - \phi_1^*)^2 + \phi_0 \phi_2 + \phi_0^* \phi_2^*$ and
$(\tilde F \tilde F)_{m \bar m} =  (\tilde F \tilde F)_{\bar m m}
= (\phi_1 + \phi_1^*)^2 - \phi_0 \phi_2 - \phi_0^* \phi_2^*$.

\vskip0.3cm
The Lorentz invariants are 
\begin{eqnarray}
\FF = {1\over4}F_{\m\n} F^{\m\n} ~=~ -{1\over2}({\bf E}^2 - {\bf B}^2)
~=~ 2~ \Rea (\phi_0 \phi_2 - \phi_1^2) \nonumber \\
\GG = {1\over4}F_{\m\n} \tilde F^{\m\n} ~=~ {\bf E}.{\bf B}  
~=~ 2~ \Ima (\phi_0 \phi_2 - \phi_1^2)~~~~~~~~~~~~~~ 
\label{eq:hi} 
\end{eqnarray}
and so $\XX = {1\over2}(\FF + i\GG) = \phi_0 \phi_2 - \phi_1^2$.
We also note the useful identity
\begin{equation}
(\tilde F \tilde F)_{\m\n} - (F F)_{\m\n} = 2 \FF g_{\m\n}
\label{eq:hff}
\end{equation}

\vskip0.3cm
The energy-momentum tensor is defined in pure Maxwell electrodynamics as
\begin{equation}
T_{\m\n} ~=~ - F_{\m\l} F_\n{}^\l + {1\over4} g_{\m\n} F^2  
\label{eq:hj}
\end{equation}
In terms of the NP field strength scalars, this can be written as
\begin{eqnarray}
T_{\m\n} ~=~ 2\Bigl( \phi_1 \phi_1^* (\ell_\m n_\n) + \phi_1 \phi_1^*
(m_\m \bar m_\n) - \phi_2 \phi_1^* (\ell_\m m_\n) - \phi_2^* \phi_1 
(\ell_\m \bar m_\n)  - \phi_0^* \phi_1 (n_\m m_\n) ~~~~\nonumber \\
- \phi_0 \phi_1^* (n_\m \bar m_\n) 
+ \phi_2 \phi_2^* \ell_\m \ell_\n + \phi_0 \phi_0^* n_\m n_\n
+ \phi_2 \phi_0^* m_\m m_\n + \phi_2^* \phi_0 \bar m_\m \bar m_\n \Bigr)
\nonumber \\
\label{eq:hk}   
\end{eqnarray}
or rather more concisely,
\begin{equation}
T_{AB} ~=~ 2~\left(\matrix{\phi_0 \phi_0^* & \phi_1 \phi_1^* 
&\phi_0 \phi_1^* & \phi_1 \phi_0^* \cr
\phi_1 \phi_1^* & \phi_2 \phi_2^* & \phi_1 \phi_2^* & \phi_2 \phi_1^* \cr
\phi_0 \phi_1^* & \phi_1 \phi_2^* & \phi_0 \phi_2^* & \phi_1 \phi_1^* \cr
\phi_1 \phi_0^* & \phi_2 \phi_1^* & \phi_1 \phi_1^* & \phi_2\phi_0^* \cr}
\right)
\label{eq:hl}  
\end{equation}
This factorisation property is the origin of many of the simplifications 
which follow from the NP formalism for electrodynamics.

\vfill\eject

\end{document}